\documentclass{aa} 
\usepackage{natbib}
\usepackage{graphicx}
\usepackage{txfonts}
\usepackage{enumitem}

\begin{document} 
\title{The evolution of the dust temperatures of galaxies in the SFR$-M_{\ast}$ plane up to $z$$\,\thicksim\,$$2$\thanks{\textit{Herschel} is an ESA space observatory with science instruments provided by European-led Principal Investigator consortia and with important participation from NASA.}}
\author{B.~Magnelli\inst{1,2}
	\and
	D.~Lutz\inst{1}      
	\and
	A.~Saintonge\inst{1,3}
	\and
      	S.~Berta\inst{1}
         \and
	P.~Santini\inst{4}
	\and
	M.~Symeonidis\inst{5,6}
	\and
	B.~Altieri\inst{7}
	\and
	P.~Andreani\inst{8,9}
	\and
	H.~Aussel\inst{10}
	\and
	M.~B\'ethermin\inst{10}
	\and
	J.~Bock\inst{11,12}
	\and
	A.~Bongiovanni\inst{13,14}
	\and
	J.~Cepa\inst{13,14}
	\and
	A.~Cimatti\inst{15}
	\and
	A.~Conley\inst{16}
	\and
	E.~Daddi\inst{10}
	\and
	D.~Elbaz\inst{10}
	\and
	N.~M.~F{\"o}rster~Schreiber\inst{1}
	\and
	R.~Genzel\inst{1}
	\and
	R.~J.~Ivison\inst{17}
	\and
	E.~Le Floc'h\inst{10}
	\and
	G.~Magdis\inst{18}
	\and
	R.~Maiolino\inst{4}
	\and
	R.~Nordon\inst{19}
	\and
	S.~J.~Oliver\inst{5}
	\and
	M.~Page\inst{6}
	\and
	A.~P{\'e}rez~Garc{\'\i}a\inst{13,14}
	\and
	A.~Poglitsch\inst{1}
	\and
	P.~Popesso\inst{1}
	\and
	F.~Pozzi\inst{15}
	\and
	L.~Riguccini\inst{20}
	\and
	G.~Rodighiero\inst{21}
	\and
	D.~Rosario\inst{1}
	\and
	I.~Roseboom\inst{17}
	\and
	M.~Sanchez-Portal\inst{7}
	\and
	D.~Scott\inst{22}
	\and
	E.~Sturm\inst{1}
	\and
	L.~J.~Tacconi\inst{1}
	\and
	I.~Valtchanov\inst{7}
	\and
	L.~Wang\inst{5}
	\and
	S.~Wuyts\inst{1}
        }
\institute{
Max-Planck-Institut f\"{u}r extraterrestrische Physik, Postfach 1312, Giessenbachstra\ss e 1, 85741 Garching, Germany
\and
Argelander Institut f\"ur Astronomy, Bonn University, Auf dem H\"ugel 71, D-53121 Bonn, Germany\\  \email{magnelli@astro.uni-bonn.de}
\and
Department of Physics and Astronomy, University College London, Gower Street, London, WC1E 6BT, UK
\and
INAF - Osservatorio Astronomico di Roma, via di Frascati 33, 00040 Monte Porzio Catone, Italy 
\and
Astronomy Centre, Dept. of Physics \& Astronomy, University of Sussex, Brighton BN1 9QH, UK 
\and
Mullard Space Science Laboratory, University College London, Holmbury St Mary, Dorking, Surrey RH5 6NT, UK
\and
Herschel Science Centre, ESAC, Villanueva de la Ca\~nada, 28691 Madrid, Spain 
\and
ESO, Karl-Schwarzschild-Stra\ss e 2, 85748 Garching, Germany 
\and
INAF - Osservatorio Astronomico di Trieste, via Tiepolo 11, 34143 Trieste, Italy 
\begin{center}\vskip 0pt \small \it (See Appendix \ref{sect:affiliations} for the remaining affiliations)\end{center}
}
\date{Received ??; accepted ??}
\abstract{We study the evolution of the dust temperature of galaxies in the SFR$-M_{\ast}$ plane up to $z\thicksim2$ using far-infrared and submillimetre observations from the \textit{Herschel} Space Observatory taken as part of the PACS Evolutionary Probe (PEP) and \textit{Herschel} Multi-tiered Extragalactic Survey (HerMES) guaranteed time key programmes.
Starting from a sample of galaxies with reliable star-formation rates (SFRs), stellar masses ($M_{\ast}$) and redshift estimates, we grid the SFR$-M_{\ast}$ parameter space in several redshift ranges and estimate the mean dust temperature ($T_{{\rm dust}}$) of each SFR$-$$M_{\ast}$$-$$z$ bin.
Dust temperatures are inferred using the stacked far-infrared flux densities (100$-$500$\,\mu$m) of our SFR$-$$M_{\ast}$$-$$z$ bins.
At all redshifts, the dust temperature of galaxies smoothly increases with rest-frame infrared luminosities ($L_{\rm IR}$), specific SFRs (SSFR; i.e., SFR/$M_{\ast}$) and distances with respect to the main sequence (MS) of the SFR$-M_{\ast}$ plane (i.e., ${\rm \Delta log(SSFR)_{\rm MS}=log[SSFR(galaxy)/SSFR_{\rm MS}}(M_{\ast},z)]$).
The $T_{{\rm dust}}$$-$${\rm SSFR}$ and $T_{{\rm dust}}$$\,-\,$${\rm \Delta log(SSFR)_{\rm MS}}$ correlations are statistically much more significant than the $T_{{\rm dust}}$$-$$L_{\rm IR}$ one.
While the slopes of these three correlations are redshift-independent, their normalizations evolve smoothly from $z=0$ and $z\thicksim2$.
We convert these results into a recipe to derive $T_{\rm dust}$ from SFR, $M_{\ast}$ and $z$, valid out to $z$$\,\thicksim\,$$2$ and for the stellar mass and SFR range covered by our stacking analysis.
The existence of a strong $T_{{\rm dust}}$$-$${\rm \Delta log(SSFR)_{\rm MS}}$ correlation provides us with several pieces of information on the dust and gas content of galaxies.
Firstly, the slope of the $T_{{\rm dust}}$$-$${\rm \Delta log(SSFR)_{\rm MS}}$ correlation can be explained by the increase of the star-formation efficiency (SFE; SFR/$M_{\rm gas}$) with $\Delta$log$({\rm SSFR})_{\rm MS}$ as found locally by molecular gas studies.
Secondly, at fixed $\Delta$log$({\rm SSFR})_{\rm MS}$, the constant dust temperature observed in galaxies probing large ranges in SFR and $M_{\ast}$ can be explained by an increase or decrease of the number of star-forming regions with comparable SFE enclosed in them.
And thirdly, at high redshift, the normalization towards hotter dust temperature of the $T_{{\rm dust}}$$-$${\rm \Delta log(SSFR)_{\rm MS}}$ correlation can be explained by the decrease of the metallicities of galaxies or by the increase of the SFE of MS galaxies.
All these results support the hypothesis that the conditions prevailing in the star-forming regions of MS and far-above-MS galaxies are different.
MS galaxies have star-forming regions with low SFEs and thus cold dust, while galaxies situated far above the MS seem to be in a starbursting phase characterized by star-forming regions with high SFEs and thus hot dust.}
\authorrunning{Magnelli et al. }
\titlerunning{Dust temperature properties of galaxies in the SFR$-M_{\ast}$ plane}
\maketitle
\section{Introduction}
Over the last $10\,$Gyr of lookback time, we observe a clear correlation between the star formation rate (SFR) and the stellar mass ($M_{\ast}$) of star-forming galaxies, SFR$\,\propto$$\,M_{\ast}^\alpha$, with $0.5<\alpha<1.0$ \citep[][]{brinchmann_2004,schiminovich_2007,noeske_2007a,elbaz_2007,daddi_2007a,pannella_2009,dunne_2009,rodighiero_2010b,oliver_2010,magdis_2010b,karim_2011,mancini_2011,whitaker_2012}.
The existence of this so-called ``main sequence'' (MS) of star-formation, whose normalization declines from $z=2.5$ to 0, is usually taken as evidence that the bulk of the star-forming galaxy (SFG) population is evolving through a steady mode of star-formation, likely sustained by the accretion of cold gas from the intergalactic medium (IGM) or along the cosmic web \citep[e.g.,][]{dekel_2009}.
In that picture, occasional major merger events create extreme systems with intense short-lived starbursts, which are offset from the MS of star-formation \citep[see e.g.,][]{engel_2010}.
One strong observational confirmation of this interpretation is that the physical properties of star-forming galaxies are fundamentally linked to their positions with respect to the MS of star-formation (i.e., ${\rm \Delta log(SSFR)_{\rm MS}=log[SSFR(galaxy)/SSFR_{\rm MS}}(M_{\ast},z)]$).
On the one hand, galaxies situated on the MS of star-formation typically have a disk-like morphology, relatively low SFR surface density \citep{wuyts_2011b}, large ratio of mid-infrared polycyclic aromatic hydrocarbon (PAHs) to far-infrared (FIR) emission \citep[][]{elbaz_2011,nordon_2012} and high \ion{C}{II}-to-FIR ratio \citep{gracia_carpio_2011}.
On the other hand, off-MS galaxies (throughout, the term ``off-MS galaxies'' refers to galaxies situated \textit{above} the MS) have cuspier morphology, higher SFR surface density, smaller PAH-to-FIR ratio and smaller \ion{C}{II}-to-FIR ratio.
Additionally, on- and off-MS galaxies also exhibit fundamental differences in the physical conditions prevailing in their giant molecular clouds (GMCs).
MS galaxies have high CO-to-H$_{2}$ conversion factors, consistent with star-forming regions mainly constituted by well virialized GMCs, while off-MS galaxies have low CO-to-H$_{2}$ conversion factors, consistent with star-forming regions being constituted by un-virialized GMCs, as observed in local major-mergers \citep{magnelli_2012b,genzel_2010,daddi_2010,magdis_2012}.

The task at hand is now to confirm this interpretation of on- and above-MS galaxies,  but also to understand the evolutionary scenario linking this population of star-forming galaxies with the large population of massive (i.e., $M_{\ast}$$\,>\,$$10^{10.5}\,$M$_{\odot}$) and passive (i.e., SFR$\,<\,$1\,M$_{\odot}\,$yr$^{-1}$) galaxies constituting the ``red sequence'' of the optical colour-magnitude diagram \citep[e.g.,][]{baldry_2004,faber_2007}, and which exists over the full redshift range covered by our study \citep{fontana_2009}.

In this context, we study here the evolution of the dust temperature ($T_{\rm dust}$) of galaxies in the SFR$-M_{\ast}$ plane up to $z$$\,\thicksim\,$$2$.
Such an analysis is crucial, because estimates of $T_{\rm dust}$ provide us with information on the physical conditions prevailing in the star-forming regions of galaxies.
In particular, because there is a link between $T_{\rm dust}$,  the radiation field, dust and gas content (via the gas-to-dust ratio), the existence of a positive correlation between dust temperature and ${\rm \Delta log(SSFR)_{\rm MS}}$ would give strong support to the usual interpretation of the MS of star-formation.
Indeed, MS galaxies with their low SFR surface densities and star-formation efficiencies (SFE; SFR/$M_{\rm gas}$; Saintonge et al. \citeyear{saintonge_2012}) should have relatively low dust temperatures, while off-MS galaxies with their high SFR surface densities and SFEs should have hotter dust temperatures.
In addition, we note that studying variations of the dust temperature in the SFR$-M_{\ast}$ plane is also important to test the accuracy of far-infrared spectral energy distribution (SED) libraries routinely used to infer total infrared luminosities from monochromatic observations, or used in backward evolutionary models \citep[see e.g.,][]{valiante_2009,bethermin_2012b}.

Estimating $T_{\rm dust}$ for a large sample of galaxies up to $z\thicksim2$ was only made possible by the \textit{Herschel} Space Observatory \citep{pilbratt_2010}.  With the Photodetector Array Camera and Spectrometer \citep[PACS;][]{poglitsch_2010} instrument and the Spectral and Photometric Imaging REceiver \citep[SPIRE;][]{griffin_2010}, it is possible to probe the peak of the rest frame far-infrared/submillimetre emission of galaxies.
Here, we use the deepest PACS and SPIRE observations obtained as part of the PACS Evolutionary Probe \citep[PEP\footnote{http://www.mpe.mpg.de/ir/Research/PEP};][]{lutz_2011} and the \textit{Herschel} Multi-tiered Extragalactic Survey \citep[HerMES\footnote{http://hermes.sussex.ac.uk};][]{oliver_2012} guaranteed time key programmes covering an area of $\thicksim\,$$2.0\,$deg$^{2}$.
The combination of these PACS and SPIRE observations, together with a careful stacking analysis, provide a very powerful tool to study the far-infrared/submillimetre SEDs of high-redshift galaxies. 
In particular, we are able to grid the SFR-$M_{\ast }$ plane in several redshift bins, extending up to $z\thicksim2$, and to estimate the dust temperature of each SFR$-$$M_{\ast}$$-$$z$ bin.
Based on these estimates, we study the correlation between the dust temperature of galaxies and their infrared luminosities \citep[see also][]{symeonidis_2013}, specific SFRs (SSFR; i.e., SFR/$M_{\ast}$) and their distances with respect to MS of star-formation.
Then, using a simple model to link $T_{\rm dust}$ with the dust and gas content of galaxies, we ascribe these correlations to variations in the physical conditions prevailing in their  star-forming regions.

The paper is structured as follows.
In section \ref{sec:data} we present the \textit{Herschel} data and the multi-wavelength sample used in our study.
Section \ref{sec:dust analysis} presents the method used to infer the evolution of the dust temperature in the SFR$-M_{\ast}$ plane from individual detections and a careful stacking analysis.
We discuss the selection function of our sample in Section \ref{subsec:parameter space} and present the variation of the dust temperature in the SFR$-M_{\ast}$ plane in Section \ref{subsec:tdust sfr-mstar}. 
Results\footnote{A simple IDL code implementing our results is available from the author at \texttt{www.mpe.mpg.de/ir/Research/PEP/Tdust\_sSFR}. This code predicts the dust temperature and the FIR/mm flux densities of a galaxy from its SFR, $M_{\ast}$ and $z$.} are discussed in Section \ref{sec:discussion}.
Finally, we summarize our findings in Section \ref{sec:conclusion}.

Throughout the paper we use a cosmology with $H_{0}=71\,\rm{km\,s^{-1}\,Mpc^{-1}}$, $\Omega_{\Lambda}=0.73$ and $\Omega_{\rm M}=0.27$.
A \citet{chabrier_2003} initial mass function (IMF) is always assumed. 
\section{Data\label{sec:data}}
\subsection{Herschel observations\label{sec:herschel}}

In order to derive the dust temperature of our galaxies, we use deep PACS 100 and 160$\,\mu$m and SPIRE 250, 350 and 500$\,\mu$m observations provided by the \textit{Herschel} Space Observatory.
PACS observations were taken as part of the PACS Evolutionary Probe \citep[PEP;][]{lutz_2011} guaranteed time key programme, while the SPIRE observations were taken as part of the \textit{Herschel} Multi-tiered Extragalactic Survey \citep[HerMES;][]{oliver_2012}.
The combination of these two sets of observations provides a unique and powerful tool to study the far-infrared SEDs of galaxies.
The PEP and HerMES surveys and data reduction methods are described in \citet{lutz_2011} and \citet{oliver_2012}, respectively.
Here, we only summarise the properties relevant for our study.

From the PEP and HerMES programmes, we used the observations of the Great Observatories Origins Deep Survey-North (GOODS-N) and -South (GOODS-S) fields and the Cosmological evolution survey (COSMOS) field.
\textit{Herschel} flux densities were derived with a point-spread-function-fitting analysis, guided using the position of sources detected in deep 24~$\mu$m observations from the Multiband Imaging Photometer \citep[MIPS;][]{rieke_2004} onboard the \textit{Spitzer} Space Observatory.
This method has the advantage that it deals in large part with the blending issues encountered in dense fields and provides a straightforward association between MIPS, PACS and SPIRE sources as well as with the IRAC (Infrared Array Camera) sources from which the MIPS-24$\,\mu$m catalogues were constructed \citep{magnelli_2011a,lefloch_2009}.
This MIPS-24$\,\mu$m-guided extraction is reliable because it has been shown that, even in the deepest PACS/SPIRE field, our MIPS-24$\,\mu$m catalogues are deep enough to contain almost all PACS/SPIRE sources \citep{magdis_2011,lutz_2011,roseboom_2010,bethermin_2012}.

In PEP, prior source extraction was performed using the method presented in \citet{magnelli_2009}, while in HerMES it was performed using the method presented in \citet{roseboom_2010}, both consortia using consistent MIPS-24$\,\mu$m catalogues.
In GOODS-N and -S, we used the GOODS MIPS-24$\,\mu$m catalogue presented in \citet{magnelli_2009,magnelli_2011a} reaching a 3$\sigma$ limit of $20\,\mu$Jy.
In COSMOS, we used the deepest MIPS-24$\,\mu$m catalogue available, reaching a 3$\sigma$ limit of $45\,\mu$Jy \citep{lefloch_2009}.
The reliability, completeness and contamination of our PACS and SPIRE catalogues were tested via Monte-Carlo simulations.
All these properties are given in \citet{berta_2011} and \citet{roseboom_2010} (see also Lutz et al., \citeyear{lutz_2011} and Oliver et al. \citeyear{oliver_2012}).
Table \ref{tab:field} summarises the depths of the PACS and SPIRE catalogs in each of the three fields.

Our IRAC-MIPS-PACS-SPIRE catalogues were cross-matched with our multi-wavelength catalogues (Section \ref{subsec: multiwavelength}), using their IRAC positions and a matching radius of 1\arcsec.
\begin{table*}
\scriptsize
\caption{\label{tab:field} Main properties of the PEP/HerMES fields used in this study.}
\centering
\begin{tabular}{ c c   ccc  ccc  ccc  ccc  ccc } 
\hline \hline
& & &\multicolumn{2}{c}{\rule{0pt}{3ex}100 $\mu$m} && \multicolumn{2}{c}{160 $\mu$m} & &\multicolumn{2}{c}{250 $\mu$m} & &\multicolumn{2}{c}{350 $\mu$m} & &\multicolumn{2}{c}{500 $\mu$m} \\
\cline{4-5} \cline{7-8} \cline{10-11} \cline{13-14} \cline{16-17}
\multicolumn{2}{c}{\rule{0pt}{2.5ex}Field}  && Eff. Area  & $3\sigma$ && Eff. Area &  $3\sigma^\mathrm{\,a}$ && Eff. Area &  $3\sigma^\mathrm{\,a}$ && Eff. Area & $3\sigma^\mathrm{\,a}$ && Eff. Area & $3\sigma^\mathrm{\,a}$ \\
 & & & arcmin$^2$  & mJy  && arcmin$^2$  & mJy  && arcmin$^2$  & mJy  && arcmin$^2$  & mJy  && arcmin$^2$  & mJy \\
\hline
\rule{0pt}{3ex}GOODS-S & ${\rm 03^{h}32^{m},\,-27^{\circ}48\arcmin}$  && 200 &  1.2 && 200 &  2.4 &&  400 &  7.8 && 400 &  9.5 && 400 &  12.1 \\ 
GOODS-N & ${\rm 12^{h}36^{m},\,+62^{\circ}14\arcmin}$ && 200 &  3.0 && 200 &  5.7 &&  900 & 9.2 && 900 & 12 && 900 & 12.1 \\ 
COSMOS & ${\rm 00^{h}00^{m},\,+02^{\circ}12\arcmin}$  && 7344 & 5.0 && 7344 &  10.2 && 7225 &  8.1 && 7225 &  10.7 && 7225 &  15.4 \\ 
\hline
\end{tabular}
\begin{list}{}{}
\item[\textbf{Notes.} ]
\item[$^{\mathrm{a}}$] In the deep 160, 250, 350 and 500$\,\mu$m observations, r.m.s values include confusion noise.
\end{list}
\end{table*}
\subsection{Multi-wavelength observations\label{subsec: multiwavelength}}

In this study we make extensive use of the large wealth of multi-wavelength observations available for the GOODS-N/S and COSMOS fields.
The full set of multi-wavelength data used in our study is described in \citet{wuyts_2011a,wuyts_2011b}.
Here, we only summarise the properties relevant for our study.

In the COSMOS field, we used the public multi-wavelength photometry described in \citet{ilbert_2009} and \citet{gabasch_2008}.
These catalogues provide photometry in a total of 36 medium and broad bands covering the SEDs from \textit{GALEX} to IRAC wavelengths.
In order to obtain reliable photometric redshifts, stellar masses and SFRs, we restricted these catalogues to $i<25$ \citep{wuyts_2011b} and to sources not flagged as problematic in the catalogue of \citet{ilbert_2009} (i.e., mostly excluding objects which are nearby bright stars).
To identify X-ray AGN we used the \textit{XMM-Newton} catalogue released by \citet{cappelluti_2009}. 
Finally, this multi-wavelength catalogue was restricted to the COSMOS area covered by our deep MIPS-PACS-SPIRE catalogue.
This leads to an effective area of 1.9 deg$^2$.

In the GOODS-S field we used the $K_{\rm s}$ ($<24.3$, 5$\sigma$) selected FIREWORKS catalogue of \citet{wuyts_2008}, which provides photometry in 16 bands from $U$ to IRAC wavelengths.
X-ray observations were taken from the \textit{Chandra} 2-Ms source catalogue of \citet{luo_2008}.
In the common region covered by our deep MIPS-PACS-SPIRE observations, our GOODS-S multi-wavelength catalogue covers an effective area of 132 arcmin$^2$. 

The GOODS-N field also benefits from extensive multi-wavelength coverage.
A $z\,$+$\,K$ selected PSF-matched catalogue was created for GOODS-N as part of the PEP survey\footnote{publicly available at http://www.mpe.mpg.de/ir/Research/PEP/} \citep{berta_2010,berta_2011}, with photometry in 16 bands from \textit{GALEX} to IRAC wavelengths, and a collection of spectroscopic redshifts \citep[mainly from][]{barger_2008}.
For uniformity with the GOODS-S catalogue, we restrict our analysis in GOODS-N to sources with $K_{\rm s}$$\,<\,$$24.3$ (3$\sigma$).
We complemented this multi-wavelength catalogue with the X-ray \textit{Chandra} 2-Ms catalogue of \citet{alexander_2003}.
In the common region covered by our deep MIPS-PACS-SPIRE observations, our GOODS-N multi-wavelength catalogue covers an effective area of 144 arcmin$^2$.

\subsection{Spectroscopic and photometric redshifts}

We use spectroscopic redshifts coming from a combination of various studies  \citep[][]{cohen_2000,wirth_2004,cowie_2004,lefevre_2004,mignoli_2005,vanzella_2006,reddy_2006b,barger_2008,cimatti_2008,lilly_2009}.
For sources with no spectroscopic redshifts, we instead use photometric redshift estimates.
Photometric redshifts were computed using \texttt{EAZY} \citep{brammer_2008} and all optical and near infrared data available.
The quality of our photometric redshifts was tested by comparing them with the spectroscopic redshifts of spectroscopically confirmed galaxies.
The median and scatter of $\Delta z/(1+z)$ are ($-0.001$; $0.015$) at $z<1.5$ and ($-0.007$; $0.052$) at $z>1.5$.
Full details are presented in \citet{wuyts_2011a,wuyts_2011b}. 

\subsection{Stellar masses\label{subsec: stellar mass}}

Stellar masses were calculated by fitting all photometric data with $\lambda_{{\rm obs}}\leqslant8\,\mu$m  to \citet{bruzual_2003} templates using \texttt{FAST} (Fitting and Assessment of Synthetic Templates; Kriek et al. \citeyear{kriek_2009}). The rest-frame template error function of \citet{brammer_2008} was also used to down-weight data points with $\lambda_{{\rm rest}}\geqslant2\,\mu$m.
In addition, we used prescriptions from \citet{wuyts_2011b} limiting the \citet{bruzual_2003} templates to models with exponentially declining star formation histories and a minimum $e$-folding time of 300$\,$Myr.
This allows for a much better agreement between SFRs derived from optical-to-near-IR SED fits and those derived using mid/far-infrared observations.  
Full details are given in \citet{wuyts_2011a,wuyts_2011b}. 
\subsection{Final sample\label{subsec:sample}}
Our final sample is $K_{\rm s}$-selected in GOODS-N ($K_{\rm s}$$\,<24.3$, down to a 3$\sigma$ significance) and GOODS-S ($K_{\rm s}$ $<24.3$, down to a 5$\sigma$ significance)  but is $i$-selected in COSMOS ($i$$\,<25$, down to a 3$\sigma$ significance).
The choice of a $i$-selected sample for COSMOS was driven by the properties of the observations publicly available for this field.
Indeed, the $K$-selected COSMOS catalogue of \citet{mccracken_2010} is relatively shallow (i.e., $K_{\rm s}$$\,<23$) and is almost fully included in the deep $i$-selected catalogue used here \citep{mccracken_2010}.

Our three multi-wavelength catalogues are not homogeneously selected and not uniform in depth, which naturally translates into different completeness limits in the SFR$-M_{\ast}$ plane.
This incompleteness could prevent us from drawing strong conclusions on the absolute number density of sources in a given SFR$-M_{\ast}$ bin and could also jeopardize the characterization of their far-infrared properties.
However, we decided here not to apply complex and very uncertain incompleteness corrections for the following reasons.

In the GOODS fields, the selection wavelength of our multi-wavelength catalogues, the $K_{\rm s}$ band, nearly translates into pure mass completeness limits.
\citet{ilbert_2013} have studied the mass completeness limits of a $K_{\rm s}$-selected sample with a magnitude limit of $K_{\rm s}$$\,=\,$$24$.
They found that, up to $z$$\,\thicksim\,$$2$, such sample was complete  down to $M_{\ast}$$\,=\,$$10^{10}\,$M$_{\odot}$.
Therefore, we conclude that with a selection limit of $K_{\rm s}$$\,<24.3$, our deep GOODS samples should provide us with a complete sample of galaxies with $M_{\ast}>10^{10}\,$M$_{\odot}$ and up to $z$$\,\thicksim\,$$2$.
In the COSMOS field, the situation is somehow different because the selection wavelength, the $i$-band, does not translate into pure mass completeness limits.
In a given SFR$-M_{\ast}$ bin, dusty systems will have fainter $i$-band magnitudes than unobscured systems, and will be more easily missed by our sample.
In this case, the source of incompleteness might correlate with the far-infrared properties of galaxies and thus bias our results.
\citet{wuyts_2011b} studied the region of the SFR-$M_{\ast}$ plane strongly affected by incompleteness in the COSMOS $i<25$ catalogue.
They used an ultradeep WFC3-selected catalog of the GOODS-S field (down to $H=27$), and looked at the fraction of galaxies in a given SFR-$M_{\ast}$ bin that had $i<25$ \citep[see Fig. 10 of][]{wuyts_2011b}.
At $M_{\ast}>10^{10}\,$M$_{\odot}$ (i.e., the mass range of interest of our study; see Section \ref{subsec:parameter space}), they found that the COSMOS $i<25$ selected catalogue is affected by incompleteness only at $z>1.5$ and that these large incompleteness mainly affect passive galaxies (completeness$\,<30\%$) and not the SFGs situated on and above the MS of star-formation (completeness$\,>60\%$).
From this analysis, we conclude that up to $z\thicksim1.5$ and for $M_{\ast}>10^{10}\,$M$_{\odot}$, incompleteness in SFR-$M_{\ast}$ bins should not bias the characterization of their far-infrared properties.
At $z>1.5$ and $M_{\ast}>10^{10}\,$M$_{\odot}$, incompleteness in the COSMOS field becomes more problematic but should not significantly affect the far-infrared properties of SFR-$M_{\ast}$ bins situated on or above the MS of star-formation.
We nonetheless verified that our results are reproduced when limiting our analysis to the GOODS fields (Appendix \ref{appendix:GOODS}).\\

Our final sample contains 8$\,$846, 4$\,$753 and 254$\,$749 sources in the GOODS-N, GOODS-S\footnote{Although the GOODS-S and -N multi-wavelength catalogues correspond both to $K_{\rm s}$ $<24.3$, the GOODS-S catalogue contains less sources than the GOODS-N catalogue because it includes sources with $>\,$5$\sigma$ while the GOODS-N multi-wavelength catalogue extends down to a 3$\sigma$ significance.} and COSMOS fields, respectively.
Of these sources, 29\%, 26\% and 3\% have a spectroscopic redshift, the rest photometric redshift estimates.
Because we study here the dust properties of galaxies, most of our results rely on the subset of sources with mid/far-infrared detections.
In GOODS-N, GOODS-S and COSMOS, 19\%, 28\% and 12\% of the galaxies have mid- or far-infrared detections, respectively.
Among those sources, 60\%, 45\% and 11\% have a spectroscopic redshift.


\subsection{Star-formation rates\label{subsec:SFR}}

We take advantage of the cross-calibrated ``ladder of SFR indicators'' established in \citet{wuyts_2011a} and used in \citet{wuyts_2011b}.
This ``ladder of SFR indicators'' has the advantage of taking the best SFR indicator available for a given galaxy, and establishes a consistent scale across all these indicators.
This ``ladder'' consists of three components: first, a step where galaxies are detected both in the rest-frame UV and the far-infrared wavelengths covered by the PACS/SPIRE \textit{Herschel} observations; second, a step where galaxies are detected both in the rest-frame UV and the mid-infrared wavelengths (i.e., in the MIPS-24$\,\mu$m passband) but are undetected in the far-infrared; third, a step where galaxies are only detected in the rest-frame UV to near infrared wavelengths.
In our final sample, $7216$, $26\,727$ and $234\,405$ sources are in the first, second and third steps of the "ladder of SFR indicators".

For galaxies with no infrared detection (i.e., galaxies of the third step), we used the SFRs estimated from the best fits obtained with \texttt{FAST} (see Section \ref{subsec: stellar mass}).
In the regime where reference SFR$_{{\rm UV+IR}}$ are available, these SED-modeled SFRs are fully consistent with those SFR$_{{\rm UV+IR}}$ estimates \citep{wuyts_2011a}.

\begin{figure*}
\center
\includegraphics[width=14.5cm]{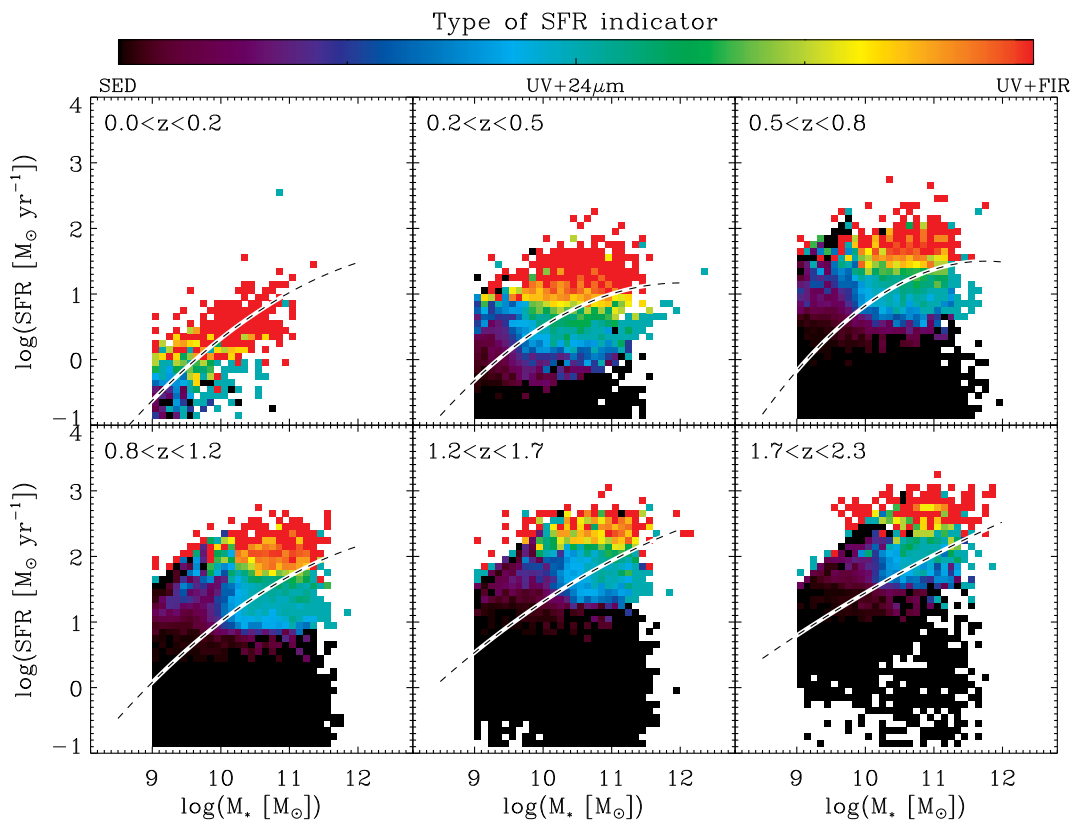}
\caption{ \label{fig:type indicators}
	Type of SFR indicator used in each SFR-$M_{\ast}$ bin.
	Red colors correspond to SFR-$M_{\ast}$ bins in which all galaxies have a far-infrared (i.e., PACS/SPIRE) detection, i.e., galaxies of the first step of the ``ladder of SFR indicator''.
	Green colors correspond to SFR-$M_{\ast}$ bins in which galaxies are not detected in the far-infrared but are detected in the mid-infrared, i.e., galaxies of the second step of the ``ladder of SFR indicator''.
	Black colors correspond to SFR-$M_{\ast}$ bins in which galaxies are only detected in the optical/near-infrared, i.e., galaxies of the third step of the ``ladder of SFR indicator''.
	Short-dashed lines on a white background show the MS of star-formation (see Sect. \ref{subsec:MS}).
}
\end{figure*}

For the first and second steps, SFRs were estimated by combining the unobscured and re-emitted emission from young stars.
This was done following \citet{kennicutt_1998} and adopting a \citet{chabrier_2003} IMF:
\begin{equation}
\label{eq: kennicutt}
{\rm SFR_{UV+IR}[M_{\odot}\,yr^{-1}]}=1.09\times10^{-10}\,(L_{{\rm IR}}[{\rm L}_{\odot}]+3.3\times L_{{\rm 2800}}[{\rm L}_{\odot}]),
\end{equation}
where $L_{2800}\equiv\nu L_{\nu}(2800\,\AA)$ is computed with \texttt{FAST} from the best-fitting SED (see Section \ref{subsec: stellar mass}) and the rest-frame infrared luminosity $L_{{\rm IR}}\equiv L(8-1000\,\mu {\rm m})$ is derived from our mid/far-infrared observations.
For galaxies with far-infrared detections, $L_{{\rm IR}}$ was inferred by fitting their far-infrared flux densities (i.e., PACS and SPIRE) with the SED template library of \citet[][DH]{dale_2002},  leaving the normalization of each SED template as a free parameter.
Examples of these fits are given in Fig. \ref{fig:sed detection}.
We note that using the SED template library of \citet{chary_2001} instead of that of DH to derive $L_{{\rm IR}}$ (again leaving the normalisation as a free parameter), has no impact on our results. 
Indeed, the $L_{{\rm IR}}^{\rm DH}$/$L_{{\rm IR}}^{\rm CE01}$ distribution has a mean value of 1 and a dispersion of 13\%.
The infrared luminosity of galaxies with only a mid-infrared detection was derived by scaling the SED template of MS galaxies \citep{elbaz_2011} to their MIPS-24$\,\mu$m flux densities.
This specific SED template was chosen because it provides good 24$\,\mu$m-to-$L_{{\rm IR}}$ conversions over a broad range of redshifts for the vast majority of star-forming galaxies  \citep[i.e., the MS galaxies;][]{elbaz_2011}.
Figure \ref{fig:type indicators} shows the localization on the SFR-$M_{\ast}$ plane of galaxies with far-infrared+UV, mid-infrared+UV and UV-only detections (i.e., galaxies from the first, second and third step of the ladder, respectively).
At $M_{\ast}>10^{10}\,$M$_{\odot}$, galaxies with only mid-infrared+UV detections are mostly situated on or below the MS of star-formation.
For these galaxies, the use of a 24$\,\mu$m-to-$L_{{\rm IR}}$ conversion based on the SED template of MS galaxies should thus be a fairly good approximation.
To further check the quality of this approximation, we verified that in all SFR-$M_{\ast}$ bins analyzed here, the mean infrared luminosities inferred from our far-infrared stacking analysis (see Section \ref{subsec:stacking}) agrees within $0.3\,$dex with those inferred from our ``ladder of SFR indicators''.

\subsection{The main sequence of star-formation\label{subsec:MS}}
\begin{figure*}
\center
\includegraphics[width=14.5cm]{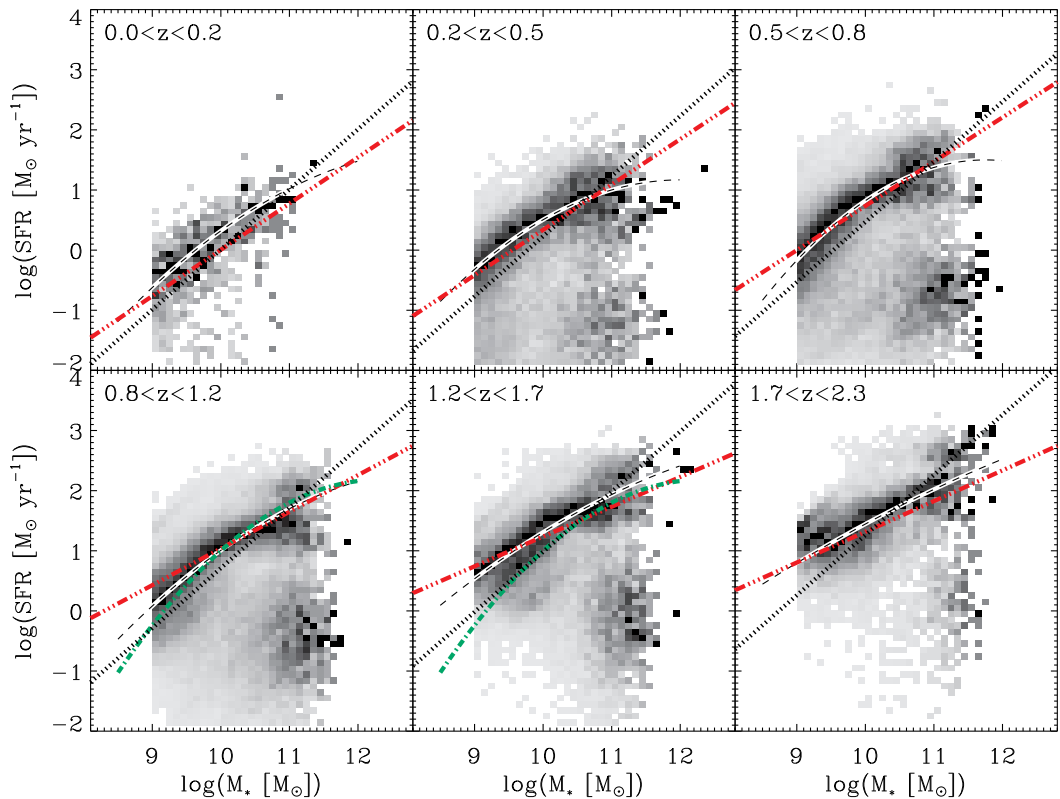}
\caption{ \label{fig: sfr mass}
Number density of sources in the SFR$-M_{\ast}$ plane.
Shading are independent for each stellar mass bin, i.e., the darkest color indicates the highest number density of sources in the stellar mass bin and not the highest number density of sources in the entire SFR$-M_{\ast}$ plane.
Short-dashed lines on a white background show the second order polynomial function fitted to MS of star-formation (see text for details).
Dotted lines present the MS found in \citet{elbaz_2011}.
The red triple-dot-dashed lines represent the MS found in \citet{rodighiero_2010b}.
The green dot-dashed lines in the $0.8<z<1.2$ and $1.2<z<1.7$ panels show the MS found in \citet{whitaker_2012} at $z\thicksim1.25$.
}
\end{figure*}

Figure \ref{fig: sfr mass} shows the number density of sources in the SFR$-M_{\ast}$ plane.
Shadings are independent for each stellar mass bin, i.e., the darkest color indicates the highest number density of sources in the stellar mass bin and not the highest number density of sources in the entire SFR$-M_{\ast}$ plane.
At $M_{\ast}$$\,<\,$$10^{11}\,$M$_{\odot}$, we observe that in a given stellar mass bin, most of the galaxies have the same SFR and this characteristic SFR increases with stellar mass.
In contrast, at $M_{\ast}$$\,>\,$$10^{11}\,$M$_{\odot}$, SFRs seems to follow a bimodal distribution; with high SFRs extending the SFR-$M_{\ast}$ correlation observed at low stellar masses, and low SFRs forming a ``cloud'' situated far below this correlation.
This bimodal distribution echoes that observed in the optical colour-magnitude diagram \citep[e.g.,][]{baldry_2004,faber_2007}, galaxies with low SFRs corresponding to galaxies from the ``red sequence'',   and galaxies with high SFRs corresponding to galaxies from the ``blue cloud''.
Focusing our attention on galaxies with relatively high SFRs, we note that over a large range of stellar masses (i.e., $10^{9}$$\,<\,$$M_{\ast}\,[{\rm M_{\odot}}]$$\,<\,$$10^{11}$), they follow a clear SFR-$M_{\ast}$ correlation.
This correlation is called ``the main sequence'' (MS) of star-formation \citep{noeske_2007a}.

At $M_{\ast}$$\,>\,$$10^{10}\,$M$_{\odot}$, there is a good agreement between the MS observed in our sample and those inferred by \citet{rodighiero_2010b} or \citet{elbaz_2011}.
However, taking into account the full range of stellar masses available, we observe a significant difference between our MS and those of \citet{rodighiero_2010b} or \citet{elbaz_2011}: the slopes of our main sequences change with stellar mass. 
Such a curved MS, not well described by a simple power-law function, has also been observed at $z$$\,\thicksim\,$$1.25$ by \citet{whitaker_2012}.
Their main sequence, fitted with a second order polynomial function, agrees with that observed in our sample.
\citet{whitaker_2012} argue that the change of the slope of the MS at high stellar masses also corresponds to a change in the intrinsic properties of the galaxies: at high stellar masses, sources are more dust rich and follow a different MS slope than lower mass galaxies with less extinction (traced by the $L_{{\rm IR}}/L_{{\rm UV}}$ ratio).

At high stellar masses ($M_{\ast}$$\,>\,$$10^{10}\,$M$_{\odot}$), where our results are focused, the slight disagreements observed between our MS and those from the literature could come from several types of uncertainties and/or selection biases \citep[see, e.g.,][]{karim_2011}.
To study all these uncertainties and selection biases is beyond the scope of this paper.
Nevertheless, because we want to investigate variations of the FIR SED properties of galaxies in the SFR$-M_{\ast}$ plane, we need to precisely define the localization of the MS of star-formation in each redshift bin.
To be self-consistent, we derived the localization of the MS using our sample.
In each redshift bin we proceeded as follows:
(i) we fitted a Gaussian function to the log(SFR) distribution of sources in each stellar mass bin.
To avoid being affected by passive galaxies, we excluded from the fit all galaxies with SFR$\,<\,$SFR$_{{\rm MS}}(M_{\ast}^{i-1})-3\times\sigma_{{\rm SFR}_{{\rm MS}}}(M_{\ast}^{i-1})$, where SFR$_{{\rm MS}}(M_{\ast}^{i-1})$  and $\sigma_{{\rm SFR}_{{\rm MS}}}(M_{\ast}^{i-1})$ are the localization and dispersion of the MS in the next lowest stellar mass bin, respectively.
(ii) The localization of the peak and the FWHM of the Gaussian function were then defined as being, respectively, the localization (i.e., SFR$_{{\rm MS}}(M_{\ast}^{i})$) and the dispersion (i.e., $\sigma_{{\rm SFR}_{{\rm MS}}}(M_{\ast}^{i})$) of the MS in this stellar mass bin.
(iii) We fitted the peak and dispersion of the MS in every stellar mass bin with a second order polynomial function.
Results of these fits are shown as short-dashed lines on a white background in Fig. \ref{fig: sfr mass} and are summarized in Table \ref{tab: MS}.
In the rest of the paper these lines are used as the MS of star-formation.
We note that the typical dispersion around these best fits is $\thicksim\,$$0.3$ dex, in agreement with previous estimates of the width of the MS of star-formation \citep[e.g.,][]{noeske_2007a}.

Because there is a good overall agreement between our fits and those from the literature at $M_{\ast}$$\,>\,$$10^{10}\,$M$_{\odot}$, the main results of our paper are essentially unchanged if using the MS defined by \citet[][see appendix \ref{appendix:MS}]{elbaz_2011}.
However, at $M_{\ast}$$\,<\,$$10^{10}\,$M$_{\odot}$, our fits significantly differ from the very steep MS defined by \citet{elbaz_2011}.
Although our results seem to be confirmed by \citet{whitaker_2012}, further investigations will be required in this stellar mass range.

\begin{table}
\caption{\label{tab: MS} Main sequence parameters}
\centering
\begin{tabular}{ c ccc} 
\hline \hline
Redshift & $A_1$ & $A_2$ & $A_3$ \\ 
\hline
$0.0<z<0.2$ & $-15.11$ & $2.27$ & $-0.073$ \\ 
$0.2<z<0.5$ & $-20.98$ & $3.66$ & $-0.152$ \\ 
$0.5<z<0.8$ & $-26.96$ & $4.77$ & $-0.199$ \\ 
$0.8<z<1.2$ & $-26.51$ & $4.77$ & $-0.202$ \\ 
$1.2<z<1.7$ & $-13.18$ & $2.18$ & $-0.073$ \\ 
$1.7<z<2.3$ & $-9.62$ & $1.60$ & $-0.050$ \\ 
\hline
\end{tabular}
\begin{list}{}{}
\item[\textbf{Notes.} ]
log(SFR$_{\rm MS})=A_{1}+A_2\times {\rm log(}M_{\ast})+A_3\times ({\rm log(}M_{\ast}))^2$
\end{list}
\end{table}
\section{Data analysis\label{sec:dust analysis}}
\subsection{Dust temperatures\label{subsec:temperature}}

In order to obtain a proxy on the dust temperature of a galaxy, one can fit its PACS/SPIRE flux densities with a single modified blackbody function using the optically thin approximation:
\begin{equation}\label{eq:BB}
S_{\nu}\propto\frac{\nu^{3+\beta}}{{\rm exp}(h\nu/kT_{{\rm dust}})-1},
\end{equation}
where $S_{\nu}$ is the flux density, $\beta$ is the dust emissivity spectral index and $T_{{\rm dust}}$ is the dust temperature.
However, we decided not to use this method for two reasons.
Firstly, a single modified blackbody model cannot fully describe the Wien side of the far-infrared SED of galaxies, because short wavelength observations are dominated by warmer or transiently heated dust components.
Therefore, one would have to exclude from this fitting procedure all flux measurements with, for example, $\lambda_{{\rm rest}}<50\,\mu$m \citep[see e.g., ][]{hwang_2010,magnelli_2012}.
At high-redshift (e.g., $z>1.0$), the exclusion of some valuable flux measurements (e.g., PACS-100$\,\mu$m) would lead to an un-optimized use of our PACS/SPIRE observations.
Secondly, while the exclusion of short wavelengths would prevent the fits from being strongly biased by hot dust components, this sharp rest-frame cut would potentially introduce a redshift dependent bias in our dust temperature estimates.
Indeed, depending on the redshift of the source, the shortest wavelength kept in the fitting procedure (i.e., with $\lambda_{{\rm rest}}>50\,\mu$m) would be close, or very close, to the wavelength cut, in which case the derived dust temperature would be affected, or strongly affected, by hotter dust components.
In the analysis done in this paper this effect would result in an artificial increase of $T_{{\rm dust}}$ with redshift of up to $\thicksim\,$$8\,$K.

For these reasons, we derived the dust temperature of a galaxy using a different approach.
(i) We assigned a dust temperature to each DH SED templates by fitting their $z=0$ simulated PACS/SPIRE flux densities with a single modified blackbody function (see Eq. \ref{eq:BB}).
(ii) We searched for the DH SED template best-fitting the observed PACS/SPIRE flux densities of the galaxy.
(iii) The dust temperature of this galaxy was then defined as being the dust temperature assigned to the corresponding DH SED template.
This method makes an optimal use of our PACS/SPIRE observations and does not introduce any redshift biases.
We also note that this method introduces a self-consistency between our dust temperature and infrared luminosity estimates for galaxies in the first step of the ``ladder of SFR indicators'' (see Sect. \ref{subsec:SFR}).
Indeed both estimates are inferred using the DH SED templates best-fitting their PACS/SPIRE flux densities.

To fit the simulated $z=0$ PACS/SPIRE flux densities of the DH SED templates with a modified blackbody function (i.e., step [i]), we fixed $\beta=1.5$ and used a standard $\chi^{2}$ minimization method.
Results of these fits are provided in Table \ref{tab:DH}.
We note that using $\beta=1.5$ systematically leads to higher dust temperatures than if using $\beta=2.0$ (i.e., $\Delta T_{{\rm dust}}\,$$\thicksim\,$$4\,$K).
This $T_{{\rm dust}}\,$$-$$\,\beta$ degeneracy further highlights the limits of a single modified blackbody model in which one has to assume, or fit, an \textit{effective} dust emissivity.
Here, we fixed $\beta=1.5$ because it will provide fair comparisons with most high-redshift studies in which the lack of (sub)mm observations does not allow clear constraints on $\beta$ and in which the \textit{effective} dust emissivity is usually fixed to 1.5 \citep[e.g.,][]{magdis_2010,chapman_2010,magnelli_2010,magnelli_2012}.
We note that leaving $\beta$ as a free parameter in our fitting procedure leads, qualitatively and quantitatively, to nearly the same $T_{{\rm dust}}$$-$$\Delta$log$({\rm SSFR})_{\rm MS}$, $T_{{\rm dust}}$$-$SSFR and $T_{{\rm dust}}$$-$$L_{\rm IR}$ relations (Section \ref{subsec:tdust sfr-mstar}).
However, due to this limitation, we stress that the absolute values of our dust temperature estimates should not be over-interpreted.

To fit the observed PACS/SPIRE flux densities of a galaxy with the DH SED templates (i.e., step [ii]), we used a standard $\chi^{2}$ minimization method.
Examples of such fits are given in Fig. \ref{fig:sed detection}.
The dust temperature of a galaxy was then defined as being the mean dust temperature assigned to the DH SED templates satisfying $\chi^2<\chi_{{\rm min}}^2+1$, and from which we also defined the 1$\sigma$ uncertainties.
This definition symmetrizes our errors bars. 
We note, however, that defining the dust temperatures as the best-fit point (i.e., where $\chi^2=\chi_{{\rm min}}^2$) does not change our results.

To ensure that our dust temperature estimates are based on reliable fits and well-sampled FIR SEDs, we used three criteria :\\
$-$ the reduced $\chi^{2}$ should be lower than 3 (i.e., typically, $\chi^{2}$$<$$6$ for N$_{\rm dof}=2$).\\
$-$ there were at least 3 PACS/SPIRE data points with S/N$>$3 to be fitted.\\
$-$ PACS/SPIRE data points with S/N$>$3 should encompass the peak of the fitted DH SED templates.\\

In the rest of the paper, we only use and discuss dust temperatures inferred from fits fulfilling these three criteria.
In Section \ref{subsec:parameter space} we will see  that this restriction does not affect our ability to study the evolution with redshift of the $T_{{\rm dust}}$$-$$L_{{\rm IR}}$, $T_{{\rm dust}}$$-$$\Delta$log$({\rm SSFR})_{\rm MS}$ and $T_{{\rm dust}}$$-$SSFR correlations.

\subsection{Stacking\label{subsec:stacking}}

In order to probe the dust temperature of galaxies below the SFR completeness limit of our PACS/SPIRE observations we used a stacking analysis.
This allows us to obtain the mean flux density of an individually undetected galaxy population by increasing their effective integration time using their stacked images.

\subsubsection{The stacking method}
Later in this paper, we demonstrate on direct detections (see Section \ref{sec:results}) that the main parameter controlling the far-infrared properties of galaxies is their localization in the SFR$-M_{\ast}$ plane. 
Consequently, the most suitable way to obtain meaningful results for our stacking analysis is to separate galaxies in different SFR$-$$M_{\ast}$ bins.
For each redshift bin, our stacking analysis was thus made as follows.
First, we gridded the SFR$-M_{\ast}$ plane.
The size of the SFR$-M_{\ast}$ grid was optimized to obtain the best balance between a large enough number of sources per grid cell to improve the noise in the stacked stamps ($\sigma_{{\rm stack}}\propto\sqrt{N}$), and a fine enough grid to meaningfully sample the SFR$-M_{\ast}$ plane.
Then, for each {\it Herschel} band (100, 160, 250, 350 and 500$\,\mu$m), we stacked all undetected galaxies in a given SFR$-M_{\ast}$ bin using the residual images\footnote{Original maps from which we removed all 3$\,\sigma$ detections}. 
Because our galaxy sample was drawn from fields with different PACS/SPIRE depths, the stacked image of each galaxy was weighted by the RMS map of our observations at the position of the source. 
Finally, the flux density was measured in each stacked image using the PSF-fitting method described in Section \ref{sec:herschel}.
The mean flux density ($F_{{\rm bin}}$) of the SFR$-M_{\ast}$ bin was then computed combining the fluxes of undetected and detected sources:
\begin{center}
\begin{equation}\label{eq:mean stack} 
S_{{\rm bin}}=\frac{m\times S_{{\rm stack}}+\sum_{i=1}^{n}S_{i}}{n+m},
\end{equation}
\end{center}
where $S_{{\rm stack}}$ is the stacked flux density of the $m$ undetected sources within the SFR$-M_{\ast}$ bin in the specific {\it Herschel} band, and  
$S_{i}$ is the flux density of the $i$-th detected source (out of a total of $n$) within the SFR$-M_{\ast}$ bin.

The uncertainty of the mean flux density of a given SFR$-M_{\ast}$ bin was computed using a standard bootstrap analysis among detections and non-detections.
Allowing for repetitions, we randomly chose $(n+m)$ sources among detections and non-detections in a given SFR$-M_{\ast}$ bin and computed their mean flux density.
This was repeated 100 times and the flux uncertainty of the SFR$-M_{\ast}$ bin was defined as the standard deviation of these 100 flux densities.
This uncertainty has the advantage of taking into account both measurement errors and the dispersion in the galaxy population.

From the mean PACS/SPIRE flux densities in each SFR$-$$M_{\ast}$ bin we derived a mean dust temperature using the same procedure as for galaxies with individual far-infrared detections and applying the same criteria to assess the accuracy of the $T_{\rm dust}$ estimates (see Section \ref{subsec:temperature}). 
In addition, we rejected dust temperature estimates in SFR-$M_{\ast}$ bins where the infrared luminosity inferred from the stacking analysis did not agree within $0.3\,$dex with the infrared luminosity inferred using our ``ladder of SFR indicators''.  
This additional criterion rejects SFR-$M_{\ast}$ bins with potentially wrong $T_{\rm dust}$ estimates and ensure a self consistency with our ``ladder of SFR indicators'' (see Sect. \ref{subsec:parameter space} and Fig. \ref{fig:completeness}). 

We note that we find similar results if we repeat the stacking analysis using the original PACS/SPIRE maps and combining all sources in a given SFR-$M_{\ast}$ bin, regardless of whether they are individually detected or not. 

\subsubsection{The effect of clustering}
While very powerful, a stacking analysis has some limitations. 
In particular, it assumes implicitly that the stacked sources are not clustered, neither within themselves (auto-correlation), nor with sources not included in the stacked sample (cross-correlation). However, this assumption is not always verified and if, for example, all stacked galaxies are situated in the vicinity of a bright infrared source, their inferred stacked flux density will be systematically overestimated.
Because our dust temperatures would be strongly affected by such biases, the effect of clustering on our stacking analysis must be investigated.

In the recent literature many techniques have been used to estimate the effect of clustering on stacked flux densities.
For example, \citet{bethermin_2012} used an approach based on simulations where the clustering effect is estimated by comparing the mean flux density measured by stacking a simulated map to the mean flux density in the corresponding mock catalogue.
Here, we adopted the same approach and evaluated the clustering effect using simulated maps of our fields.

The main principle of this method is to construct PACS and SPIRE simulated maps reproducing as well as possible the intrinsic far-infrared emission of the Universe in our fields.
To do so, we used our multi-wavelength galaxy sample which contains the positions, redshifts and SFRs of at least all galaxies with $M_{\ast}$$\,>\,$$10^{10}\,$M$_{\odot}$ and $z$$\,<\,$$2$ (see Sect. \ref{subsec:sample}).
This galaxy sample is particularly suitable because in term of intrinsic far-infrared emission, it contains sources with SFRs a factor at least 10 below the SFR thresholds equivalent to the noise level of our far-infrared images.
This ensures that simulated PACS and SPIRE maps created from this galaxy sample would contain all galaxies that can introduce some clustering biases in our FIR stacking analysis. 

From the redshift and SFR of each galaxy in our multi-wavelength sample, we inferred their simulated PACS and SPIRE flux densities using the MS galaxy SED template of \citet{elbaz_2011}.
The use of this particular SED template is appropriate because it was built to reproduce the mean FIR emission of main sequence star-forming galaxies \citep{elbaz_2011}.
Of course, real SEDs of star-forming galaxies present variations around this mean SED.
However, there are always at least 100 simulated sources in SFR-$M_{\ast}$-$z$ bins where $T_{\rm dust}$ is estimated through stacking such that the central-limit theorem applies.

From the position and simulated far-infrared flux densities of each galaxy, we then created PACS and SPIRE simulated maps using the appropriate pixel scales and PSFs.
Because the real stacking analysis was done on residual maps, galaxies with direct far-infrared detections were not introduced in these simulated ``residual'' maps. 
We note that these simulations are of course not perfect, but could be considered as the best representation we have so far of the intrinsic far-infrared properties of our fields, in particular of their clustering properties.  

Using these simulated ``residual'' maps, we estimated the effect of clustering on our stacked samples.
For each SFR$-$$M_{\ast}$$-$$z$ bin and for each PACS and SPIRE band, this was done as follows: \\
- (i) we stacked in the real residual maps the $m$ sources of this specific bin and measured the PACS and SPIRE stacked flux densities ($S_{{\rm stack}}$). \\
- (ii) we stacked in the simulated ``residual'' maps the same $m$ sources and measured their stacked simulated PACS and SPIRE flux densities ($S_{{\rm stack}}^{{\rm simu}}$). \\
- (iii) we computed, using the simulated PACS/SPIRE catalogues, the expected stacked simulated PACS and SPIRE flux densities ($S_{{\rm stack}}^{{\rm expected}}$). \\
- (iv) we compared $S_{{\rm stack}}^{{\rm simu}}$ and $S_{{\rm stack}}^{{\rm expected}}$ and if ABS(($S_{{\rm stack}}^{{\rm simu}}-S_{{\rm stack}}^{{\rm expected}}$)/$S_{{\rm stack}}^{{\rm expected}}$)$>0.5$ then the real stacked flux densities were identified as being potentially affected by clustering. This $0.5$ value was empirically defined as being the threshold above which the effect of clustering would not be captured within the flux uncertainties of our typical S/N$\,\thicksim\,$3 stacked photometries.\\

Dust temperatures estimated using one of these affected stacked flux densities were flagged and \textit{not used in the rest of our analysis}.
We adopted this very conservative approach knowing the limits of our simulations: if one band is affected by clustering, the other bands might also be affected, even if formally there are not identified as such by our simulations.
In the stellar mass range of our study, $M_{\ast}$$\,>\,$$10^{10}\,$M$_{\odot}$, and in each of our redshift bins, fewer than $10$ SFR$-$$M_{\ast}$ bins are rejected due to this clustering effect.  
Most of these SFR-$M_{\ast}$ bins are situated in the border between bins with low SFRs, and thus no stacked detections, and bins with high SFRs and thus clear stacked detections.

We stress that these tests are not perfect, but we believe that they are better than any other tests that would treat our sample in a redshift independent way.
The reliability of these tests is strengthened by the accuracy of our simulated ``residual'' maps.
Indeed, in SFR-$M_{\ast}$-$z$ bins not affected by clustering and with $S_{{\rm stack}}/N_{{\rm stack}}>3$, the log($S_{{\rm stack}}^{{\rm simu}}/S_{{\rm stack}}$) values  follow a Gaussian distribution with a median value of 0.1 dex and a dispersion of 0.1 dex.
We note that even though our simulated maps appear to be accurate, we did not use the $S_{{\rm stack}}^{{\rm simu}}/S_{{\rm stack}}^{{\rm expected}}$ ratio as a correction factor for $S_{{\rm stack}}$ because the uncertainties on these correcting factors (at least $\thicksim\,$$0.1\,$ dex from the log($S_{{\rm stack}}^{{\rm simu}}/S_{{\rm stack}}$) dispersion) would be equivalent to the flux uncertainties of our typical S/N$\,\thicksim\,$3 stacked fluxes.

\subsubsection{The effect of averaging on dust temperature}
Finally, we tested whether or not the stacking procedure used to produce the mean FIR SED of a galaxy population delivers reliable mean $T_{\rm dust}$ estimates, given that the galaxies follow a particular $L_{\rm IR}$ and $T_{\rm dust}$ distribution.  
In other words, is the mean dust temperature of a galaxy population computed from individual dust temperature measurements in agreement with the value inferred from their mean FIR SED?
To perform this test, we assumed that the FIR SEDs of a galaxy population are well described by the DH SED template library and that this galaxy population follows the local $T_{{\rm dust}}$$-$$L_{{\rm IR}}$ correlation \citep[][see also Dunne et al. \citeyear{dunne_2000}]{chapman_2003}.

\begin{figure}
\center
\includegraphics[width=9.3cm]{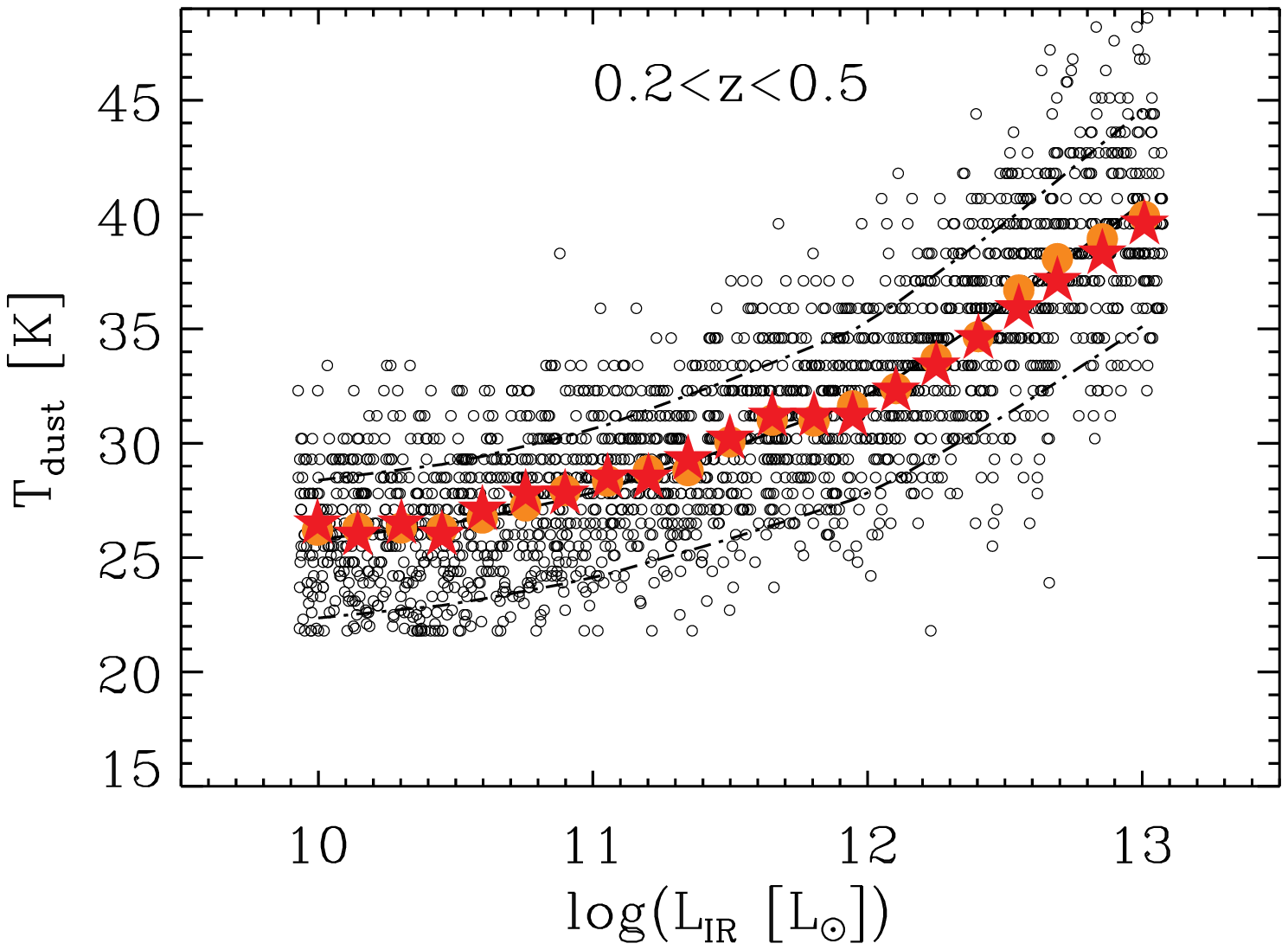}
\includegraphics[width=9.3cm]{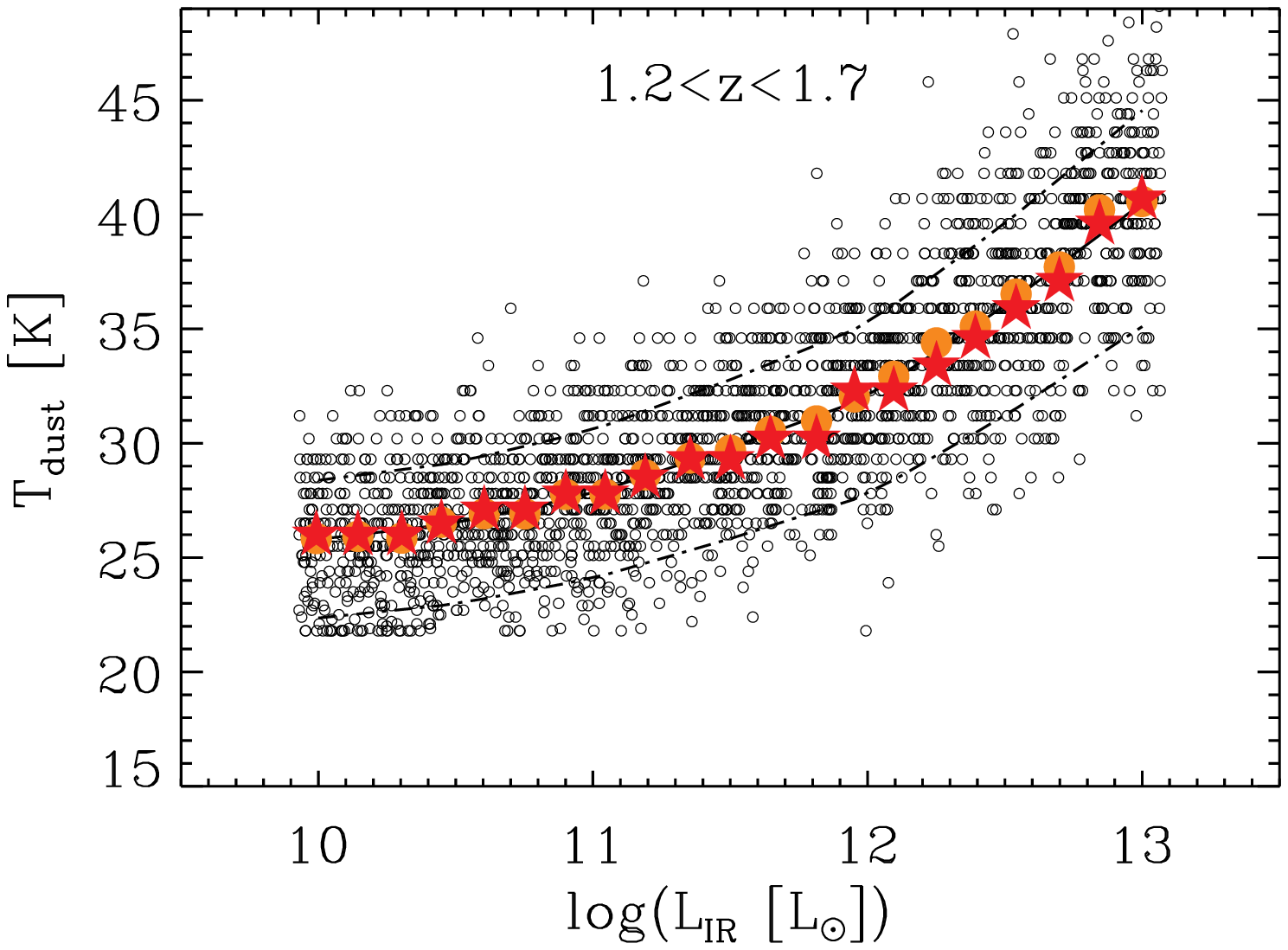}
\caption{ \label{fig:bias stack} Simulations revealing the effects of stacking on our dust temperature estimates.
Empty black circles represent our 1000 simulated galaxies following the $T_{{\rm dust}}$$-$$L_{{\rm IR}}$ correlation of \citet{chapman_2003}.
Large filled orange circles represent the mean dust temperature of galaxies in a given infrared luminosity bin.
Large filled red stars represent the dust temperature inferred from the mean PACS/SPIRE flux densities of galaxies in a given infrared luminosity bin.
The \citet{chapman_2003} derivation of the median and interquartile range of the $T_{{\rm dust}}$$-$$L_{{\rm IR}}$ relation observed at $z$$\,\thicksim\,$$0$ is shown by solid and dashed-dotted lines, linearly extrapolated to $10^{13}\,$L$_{\odot}$.
In the \textit{top} panel, simulated galaxies are in the $0.2<z<0.5$ redshift bin, while in the \textit{bottom} panel simulated galaxies are in the $1.2<z<1.7$ redshift bin.
}
\end{figure}

We first created a simulated catalogue containing 1000 galaxies. 
The redshift of each galaxy was randomly selected using a uniform redshift distribution within our redshift bins (i.e., $0<z<0.2$, $0.2<z<0.5$~...), the infrared luminosity from uniform logarithmic distribution with 10$\,<\,$log($L_{\rm IR}$[L$_{\odot}$])$\,<\,$13, and $T_{\rm dust}$ from a Gaussian distribution reproducing the local $T_{{\rm dust}}$$-$$L_{{\rm IR}}$ correlation and its dispersion \citep{chapman_2003}. Next, we assign to each simulated galaxy the appropriate DH SED template, based on our pairing between dust temperature and DH templates (see Sect. \ref{subsec:temperature}).   
Finally, simulated galaxies are separated in $L_{\rm IR}$ bins of 0.15 dex, the typical size of our SFR-$M_{\ast}$ bins, and in each $L_{\rm IR}$ bin we compute the mean $T_{\rm dust}$ value in two ways: (1) using the mean PACS/SPIRE flux densities, and (2) directly averaging the values of $T_{\rm dust}$ assigned to each galaxy.
The two sets of mean $T_{\rm dust}$ values are compared, as shown in Fig. \ref{fig:bias stack} for two redshift bins. 

We find that the mean dust temperatures and those inferred from the mean FIR SEDs only differ by a few degrees.
Moreover, we do not find any strong systematic biases as a function of  either $L_{\rm IR} $ or redshift, and therefore conclude that the dust temperatures inferred from the stacked mean FIR SEDs accurately represent the underlying galaxy population.
We note that a flattening of the $T_{{\rm dust}}$$-$$L_{{\rm IR}}$ correlation with redshift as found in \citet{symeonidis_2013} does not change our results, as they do not strongly depend on the slope of the $T_{{\rm dust}}$$-$$L_{{\rm IR}}$ correlation but rather on its dispersion.

\section{Results\label{sec:results}}
\begin{figure*}
\center
\includegraphics[width=9.1cm]{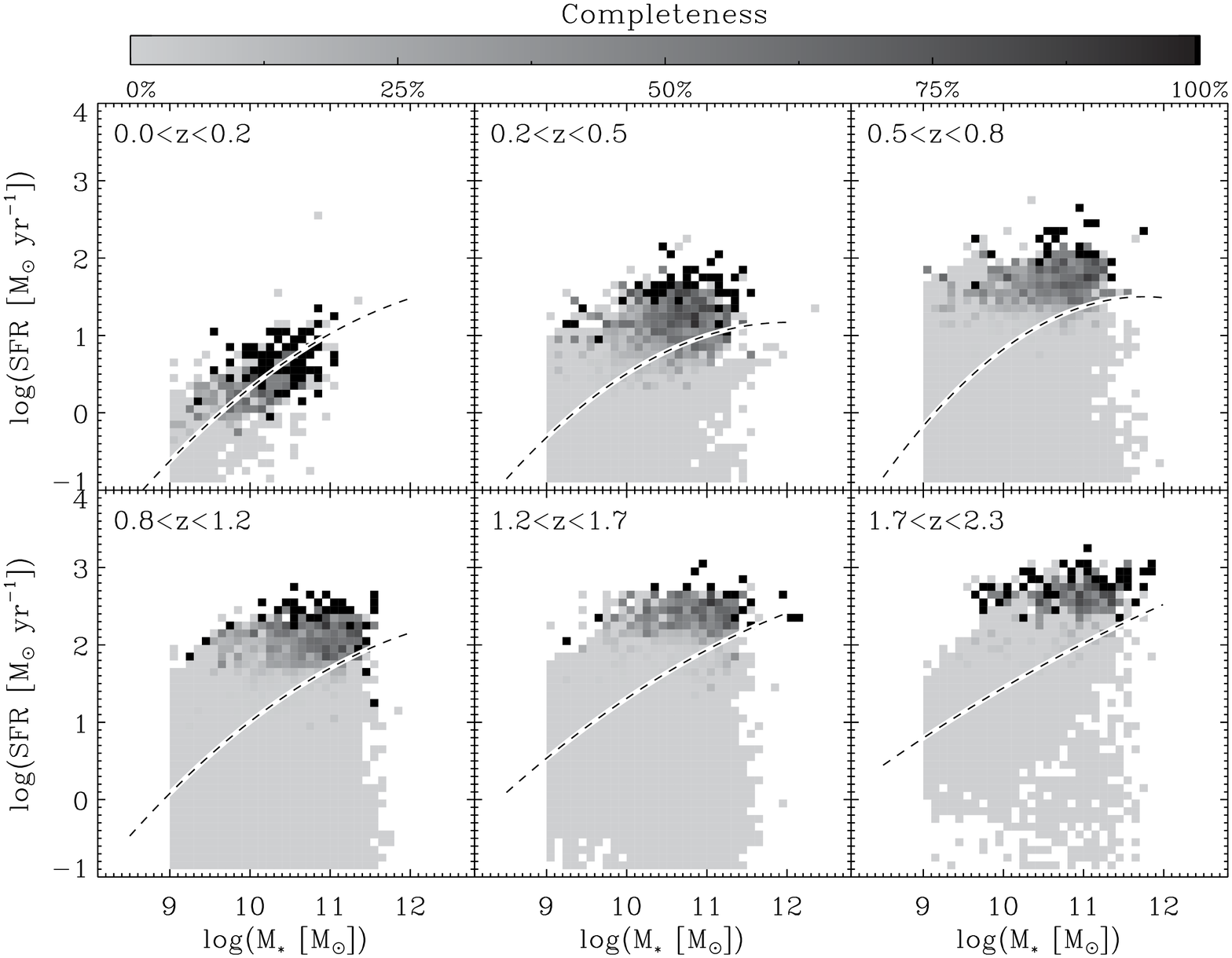}
\includegraphics[width=9.1cm]{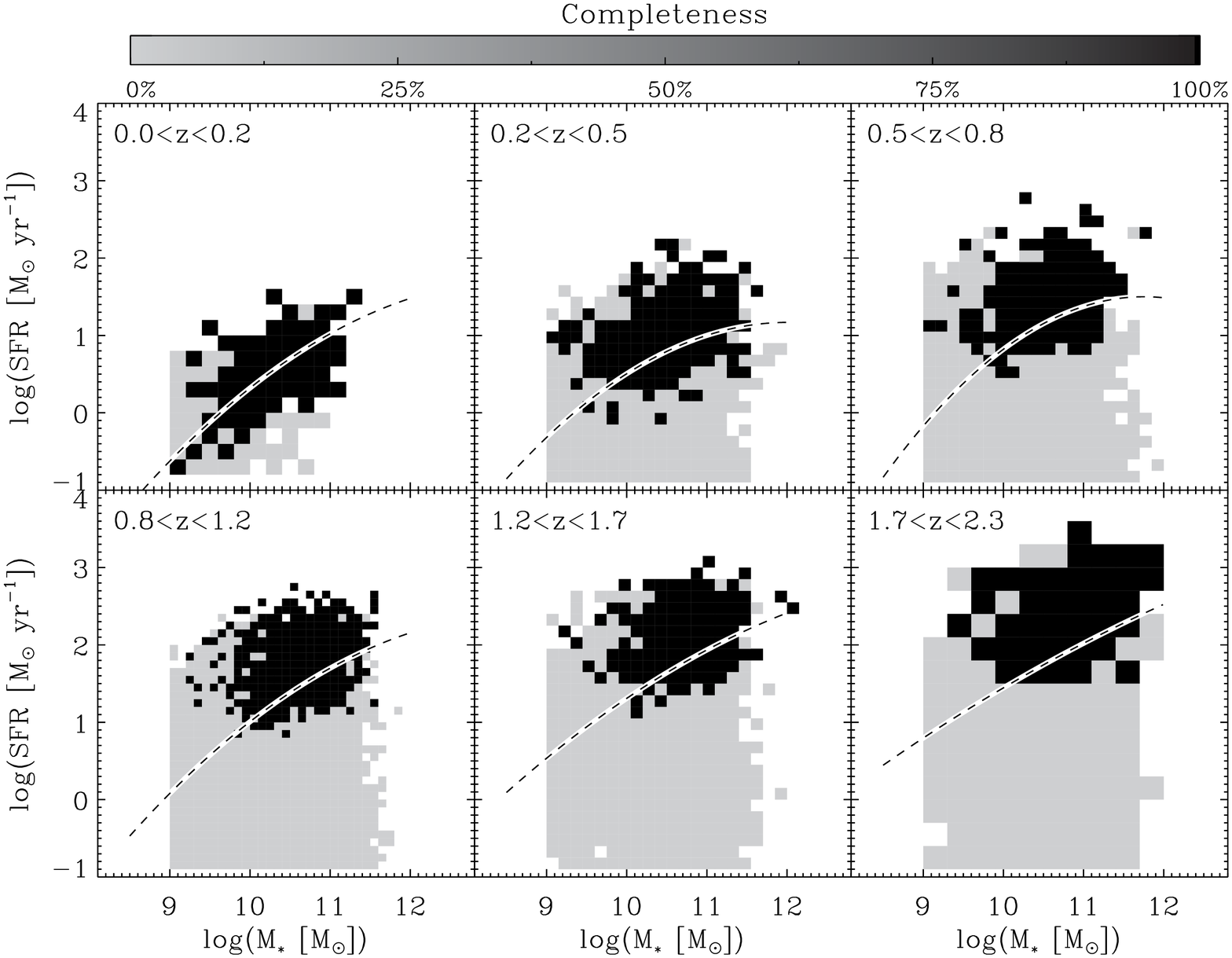}
\caption{ \label{fig:completeness}
(\textit{left}) Fraction of sources (i.e., completeness) in a given SFR-$M_{\ast}$ bin with individual PACS/SPIRE  detections and accurate dust temperature estimates.
(\textit{right}) SFR$-M_{\ast}$ bins with accurate dust temperature estimates (i.e., completeness$\,=\,$$100\%$), as inferred from our stacking analysis. 
Because each SFR-$M_{\ast}$ bin corresponds to only one set of stacked far-infrared flux densities and thus one dust temperature estimate, here the completeness can only take two different values: 0\% if the dust temperature estimate is inaccurate; 100\% if the dust temperature estimate is accurate.
The quality of our dust temperature estimates were evaluated using criteria presented in Sect. \ref{subsec:temperature} and \ref{subsec:stacking}.
Short-dashed lines on a white background show the MS of star-formation.
}
\end{figure*}
\begin{figure*}
\center
\includegraphics[width=8.9cm]{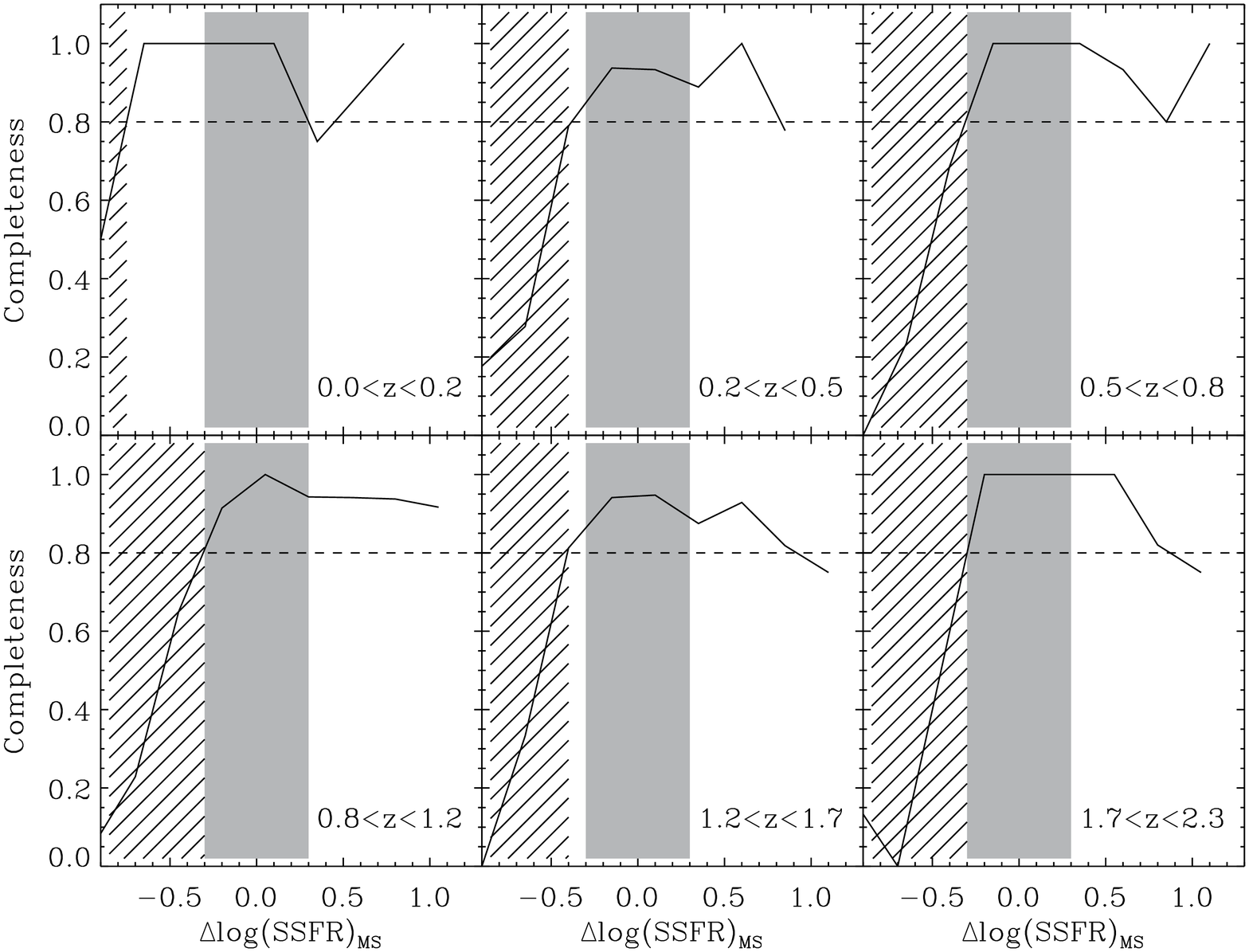}
\includegraphics[width=8.9cm]{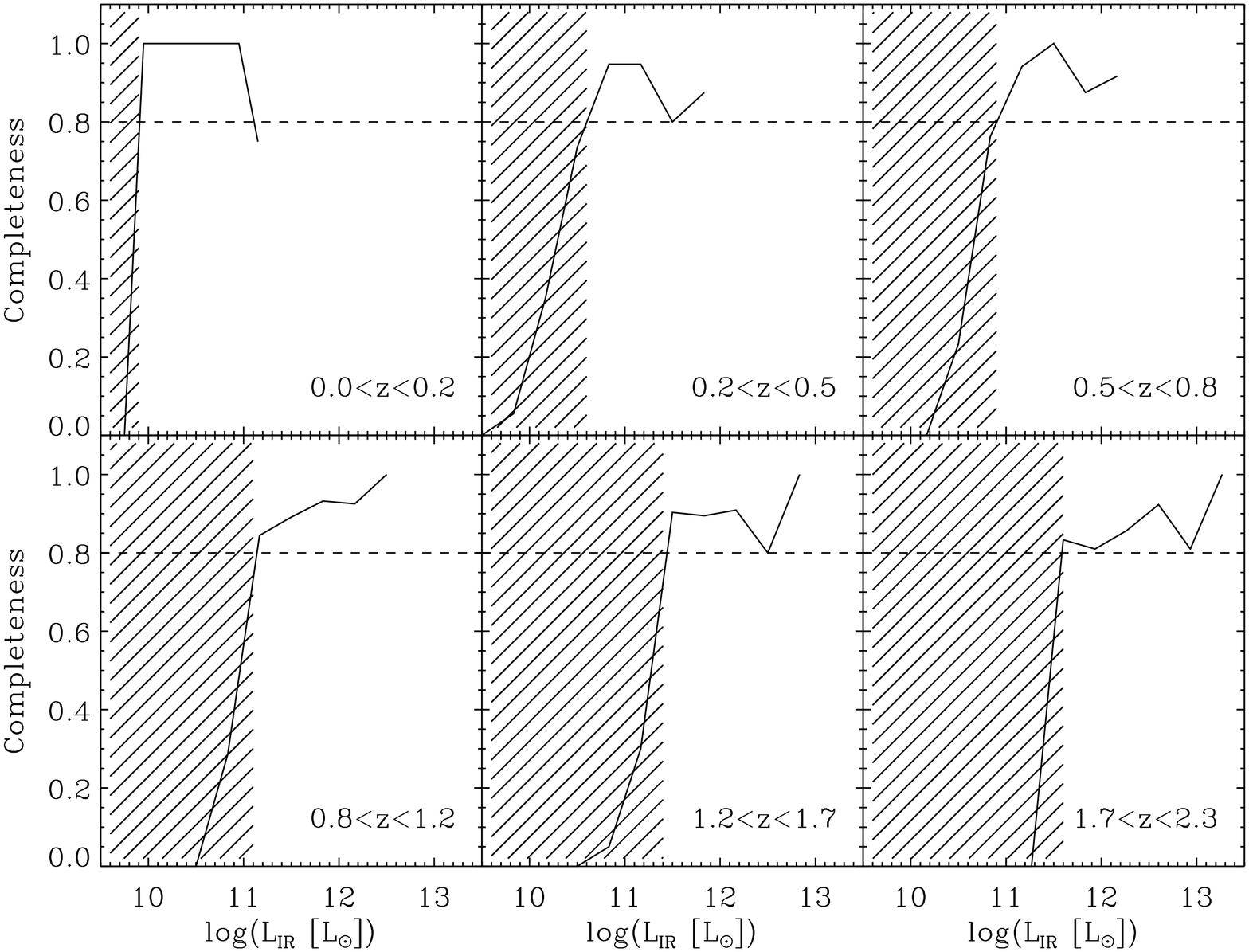}
\includegraphics[width=8.9cm]{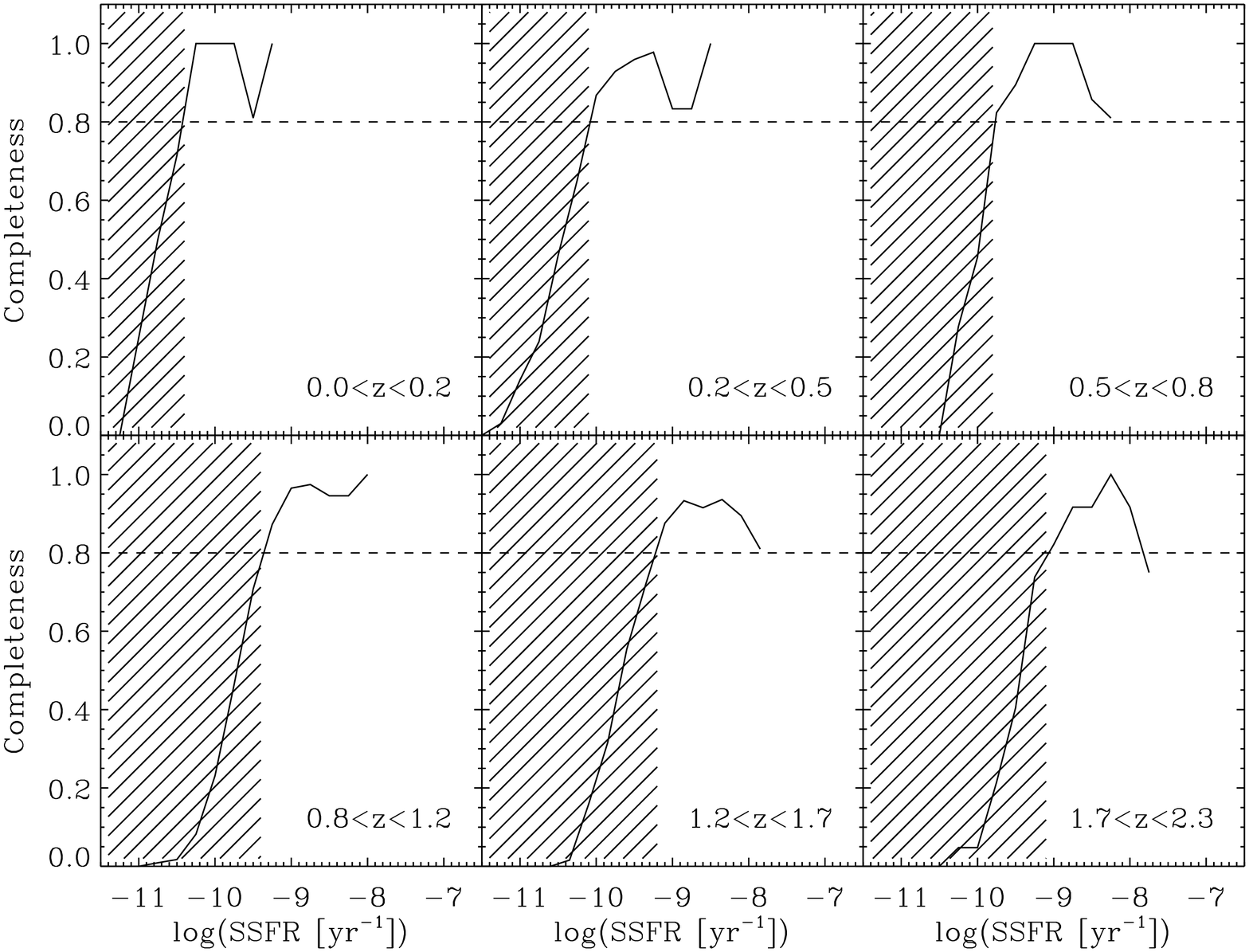}
\caption{ \label{fig:flag}
Fraction of SFR$-M_{\ast}$ bins with $M_{\ast}$$\,>\,$$10^{10}\,$M$_{\odot}$ and with accurate dust temperature estimates (see Sect.~\ref{subsec:temperature}) in our stacking analysis as function of their $\Delta$log$({\rm SSFR})_{\rm MS}$ (\textit{top left panel}), $L_{{\rm IR}}$ (\textit{top right panel}) or ${\rm SSFR}$ (\textit{bottom panel}).
Horizontal dashed lines represent the 80\% completeness limits.
Striped areas represent the regions of parameter space affected by incompleteness, i.e., where less than 80\% of our SFR$-M_{\ast}$ bins have accurate dust temperature estimates.
Shaded regions in the top left panel show the location and dispersion of the MS of star-formation.
}
\end{figure*}
\subsection{The SFR$-$$M_{\ast}$$-$$z$ parameter space\label{subsec:parameter space}}

We have seen in Section \ref{subsec:MS} that incompleteness in our initial sample should not affect our ability to study variations of $T_{\rm dust}$ in the SFR$-M_{\ast}$ plane.
However, one additional source of incompleteness might still affect our study: our ability to derive accurate $T_{\rm dust}$ for each SFR$-$$M_{\ast}$$-$$z$ bin.
Therefore, before presenting the variations of dust temperature in the SFR$-M_{\ast}$ plane, we examine which regions of the SFR$-$$M_{\ast}$$-$$z$ parameter space have accurate $T_{\rm dust}$ estimates from individual detections, and from our stacking analysis.
The left panel of Fig.~\ref{fig:completeness} presents the fraction of sources (i.e., the completeness) in a given SFR$-$$M_{\ast}$$-$$z$ bin with \textit{individual} far-infrared detections and \textit{accurate} dust temperature estimates (see Sect.~\ref{subsec:temperature}).
The right panel of Fig.~\ref{fig:completeness} presents the same quantity for the stacking analysis. 
There, each SFR-$M_{\ast}$-$z$ bin corresponds to only one set of stacked far-infrared flux densities and thus one dust temperature estimate.
Consequently, in the right panel of Fig.~\ref{fig:completeness}, the completeness can only take two different values: 0\% if the dust temperature estimate is inaccurate; 100\% if the dust temperature estimate is accurate.

For galaxies individually detected in our PACS/SPIRE images reliable dust temperatures are only obtained in SFR$-M_{\ast}$ bins situated above a given SFR threshold. 
This SFR threshold increases with redshift and translates into a threshold in $L_{\rm IR}$.
If keeping above this threshold, the   $T_{{\rm dust}}$$-$$L_{{\rm IR}}$ relation can therefore be studied without biases.
However, because the region of the SFR$-$$M_{\ast}$ plane with high completeness in all redshift bins is relatively small, we conclude that the PACS/SPIRE detections are not sufficient to draw strong conclusions on the redshift evolution of the $T_{{\rm dust}}$$-$$L_{{\rm IR}}$ relation.
The prognostic for using the PACS/SPIRE detections to study the $T_{{\rm dust}}$$-$$\Delta$log$({\rm SSFR})_{\rm MS}$ and $T_{{\rm dust}}$$-$${\rm SSFR}$ relations is even worse.
Indeed, in all but our first redshift bin, the dust temperature of MS galaxies is only constrained for those with very high stellar masses.
The stacking analysis is therefore required. 

The right panel of Fig. \ref{fig:completeness} illustrates the regions of the SFR$-$$M_{\ast}$$-$$z$ parameter space reachable with our stacking analysis.
The SFR completeness limit of the stacking analysis depends on the stellar masses: it is nearly constant at intermediate stellar masses, but increases at both very high and low  stellar masses. These variations are caused by changes in the number of galaxies in the SFR$-M_{\ast}$ bins.  At fixed SFR, the SFR$-M_{\ast}$ bins with low stellar masses are located above the MS and thus contain far fewer sources than those with intermediate stellar masses and which probe the bulk of the MS. 
Similarly, at fixed SFR, SFR$-M_{\ast}$ bins with very high stellar masses are located below the MS of star-formation and thus contain far fewer sources than those with intermediate stellar masses and which probe the MS of star-formation.  
The decrease in the number of sources in these SFR$-M_{\ast}$ bins naturally translates into higher SFR completeness limits in our stacking analysis ($\sigma_{{\rm stack}}\propto\sqrt{N_{{\rm sources}}}$).
Due to the very strong increase of the SFR completeness limits at low stellar masses, we restrict our study to galaxies with $M_{\ast}>10^{10}\,$M$_{\sun}$.

Figure \ref{fig:flag} illustrates our ability to study the variations of $T_{\rm dust}$ in the SFR$-$$M_{\ast}$ plane for galaxies with $M_{\ast}>10^{10}\,$M$_{\sun}$ through the stacking analysis.
In these figures we show the fraction of SFR$-M_{\ast}$ bins with reliable dust temperature estimates (see Sect.~\ref{subsec:temperature}) as a function of their $\Delta$log$({\rm SSFR})_{\rm MS}$, $L_{{\rm IR}}$ and ${\rm SSFR}$.
In the rest of the paper, we consider that the mean dust temperature in a given $\Delta$log$({\rm SSFR})_{\rm MS}$, $L_{{\rm IR}}$ or ${\rm SSFR}$ bin is fully constrained if and only if the completeness in this bin is at least 80\%.

In each redshift bin, our stacking analysis allows us to fully constrain the mean dust temperature of galaxies situated on and above the MS of star-formation at such a completeness level, and therefore to reliably study the redshift evolution of the $T_{{\rm dust}}$$-$$\Delta$log$({\rm SSFR})_{\rm MS}$ correlation.
The stacking analysis furthermore allows us to constrain the mean dust temperature of galaxies over more than an order of magnitude in $L_{{\rm IR}}$ and SSFR.
However, because our  $L_{{\rm IR}}$ and SSFR completeness limits increase with redshift, the $L_{{\rm IR}}$ and SSFR ranges probed in our different redshift bins do not fully overlap.  Thus while we can robustly constrain the $T_{{\rm dust}}$$-$$L_{{\rm IR}}$ and $T_{{\rm dust}}$$-$${\rm SSFR}$ correlations in all our redshift bins, the study of their evolution with redshift is somewhat limited.

\subsection{Dust temperature in the SFR$-$$M_{\ast}$$-$$z$ parameter space\label{subsec:tdust sfr-mstar}}
\begin{figure*}
\center
\includegraphics[width=13.7cm]{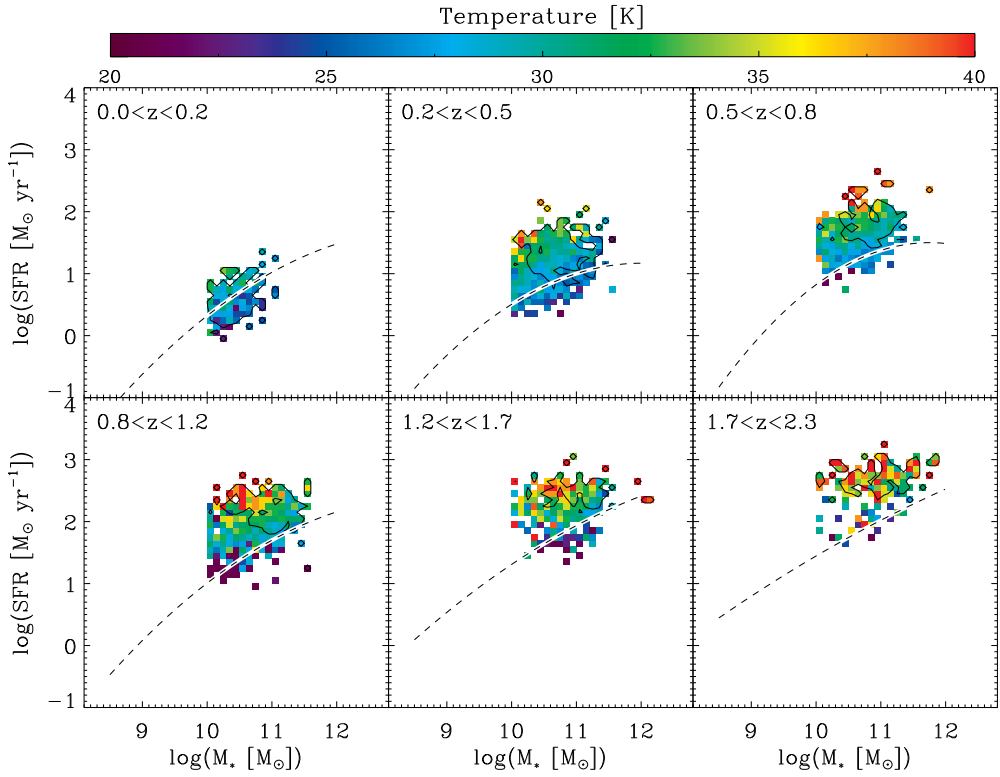}
\caption{ \label{fig:tdust}
Mean dust temperature of galaxies in the SFR$-M_{\ast}$ plane as found from individual detections.
Short-dashed lines on a white background show the MS of star-formation.
Solid contours indicate the regions in which at least 50\% of the galaxies have accurate dust temperature estimates.
Outside these regions results have to be treated with caution, because they are inferred from incomplete samples. 
Tracks of iso-dust-temperature are not characterized by vertical or horizontal lines but instead by approximately diagonal lines.
}
\end{figure*}
\begin{figure*}
\center
\includegraphics[width=13.7cm]{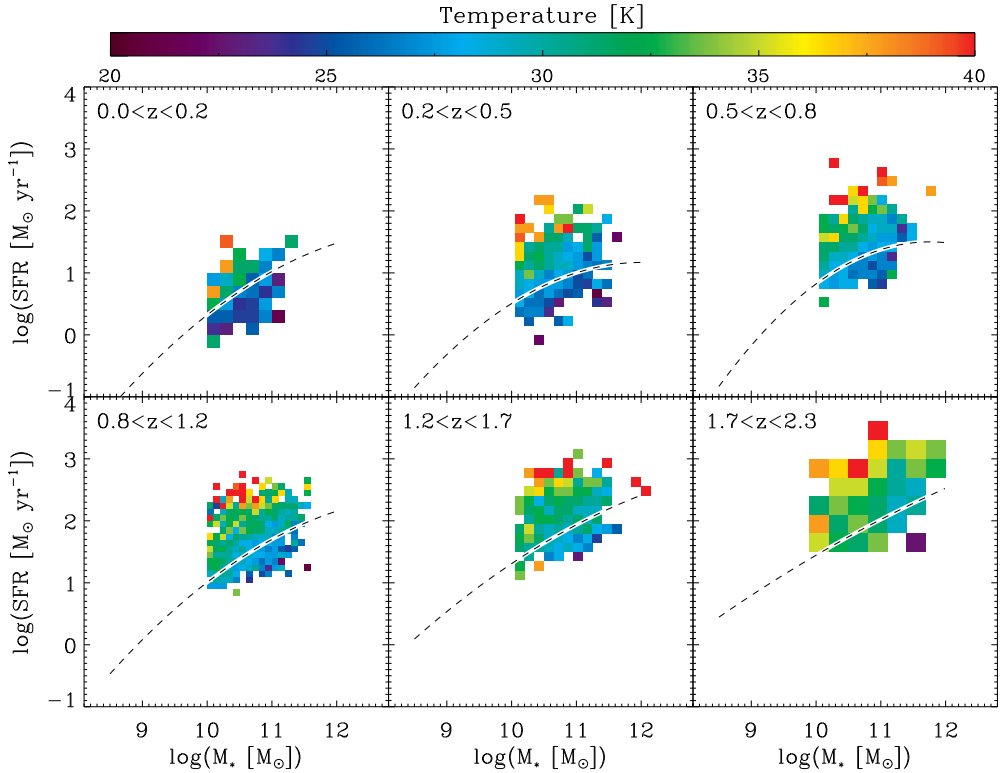}
\caption{ \label{fig:tdust stack}
Mean dust temperature of galaxies in the SFR$-M_{\ast}$ plane as found from our stacking analysis.
Short-dashed lines on a white background show the MS of star-formation.
Tracks of iso-dust-temperature are not characterized by vertical or horizontal lines but instead by approximately diagonal lines.
}
\end{figure*}

Figures \ref{fig:tdust} and \ref{fig:tdust stack} show the mean dust temperatures of galaxies in our SFR$-$$M_{\ast}$$-$$z$ bins as inferred using individual detections and our stacking analysis, respectively.
In these figures, we only used accurate dust temperature estimates (see Sect.~\ref{subsec:temperature}).
For the detections, the black contours indicate regions where at least 50\% of the galaxies in a given SFR$-M_{\ast}$ bin have reliable dust temperature estimates.
Outside these regions, results have to be treated with caution, because they rely on very incomplete samples. Both figures clearly show that in any given redshift bin, $T_{\rm dust}$ does not evolve simply with their SFR or with their stellar mass, but instead with a combination of these two parameters.  In other words, tracks of iso-dust temperature are not characterized by vertical or horizontal lines, but instead by diagonal lines.

\begin{figure*}
\center
\includegraphics[width=13.5cm]{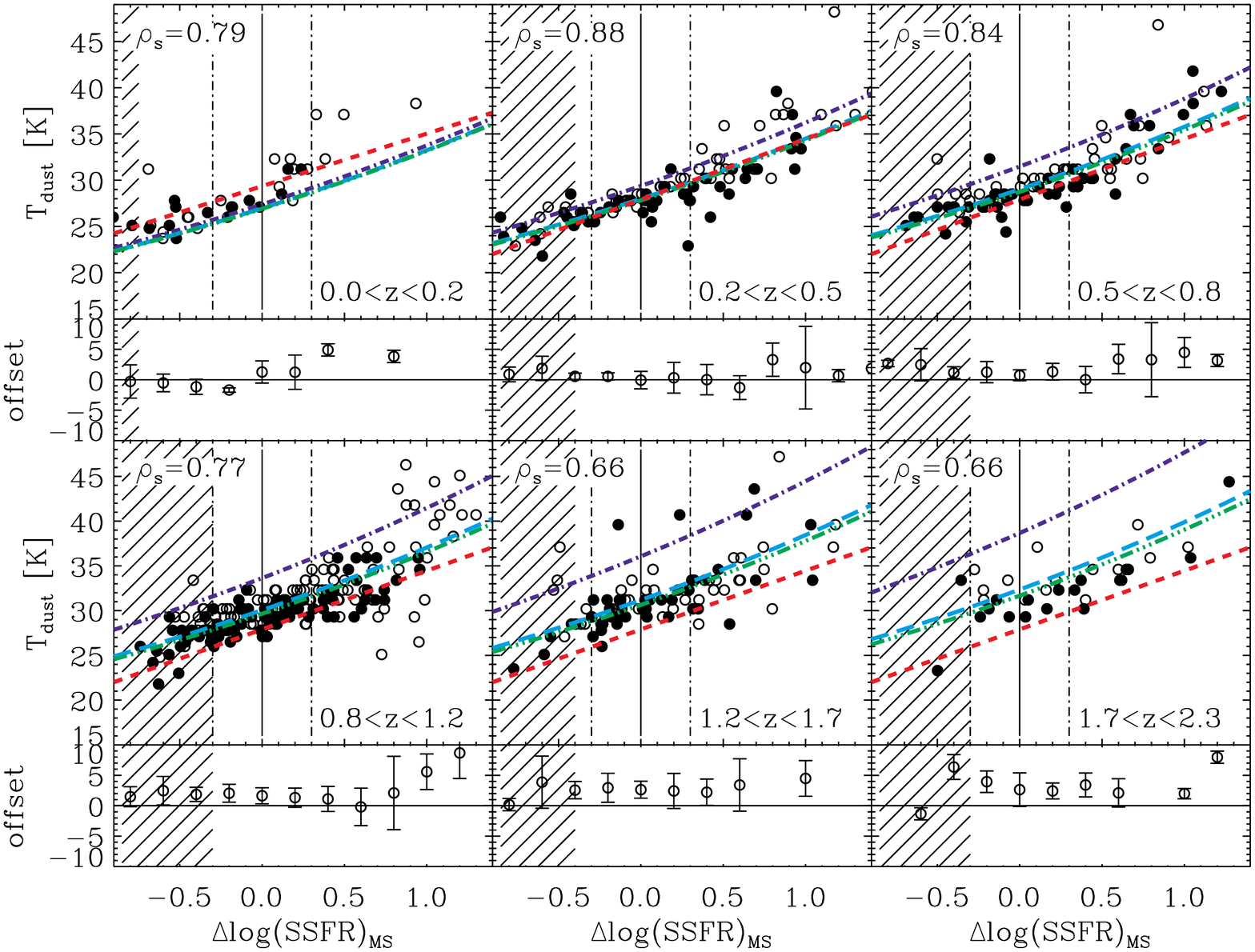}
\caption{ \label{fig:tdust dssfr}
Dust temperature of galaxies as a function of $\Delta$log$({\rm SSFR})_{\rm MS}$, as derived from our stacking analysis.
Empty circles show results for SFR$-M_{\ast}$ bins with low stellar masses, i.e., $M_{\ast}$$\,<\,$$10^{10.5}$, $10^{10.7}$, $10^{10.7}$, $10^{10.8}$, $10^{10.8}$ and $10^{10.9}\,$M$_{\odot}$ in our $0<z<0.2$, $0.2<z<0.5$, $0.5<z<0.8$, $0.8<z<1.2$, $1.2<z<1.7$ and $1.7<z<2.3$ bins, respectively.  
Filled circles show results for SFR$-M_{\ast}$ bins with high stellar masses.
In each panel, we give the Spearman correlation factor derived from our data points.
Red dashed lines correspond to a linear fit to the data points of the $0.2<z<0.5$ redshift bin.
Green triple-dot-dashed lines represent predictions inferred from the variations of the SFE with $\Delta$log$({\rm SSFR})_{\rm MS}$ and variations of the metallicity with redshift (see text for more details).
Light blue long-dashed lines represent predictions inferred from the variations of the SFE with $\Delta$log$({\rm SSFR})_{\rm MS}$ and variations of the \textit{normalization} of the SFE-$\Delta$log$({\rm SSFR})_{\rm MS}$ relation with redshift (see text for more details).
Dark blue dot-dashed lines represent predictions inferred from the variations of the SFE with $\Delta$log$({\rm SSFR})_{\rm MS}$, and variations with redshift of the normalization of the SFE-$\Delta$log$({\rm SSFR})_{\rm MS}$ relation and of the metallicity (see text for more details).
Hatched areas represented the regions of parameter space affected by incompleteness (see text and Fig.~\ref{fig:flag}).
Vertical solid and dot-dashed lines show the localization and width of the MS of star-formation.
The lower panel of each redshift bin shows the offset between the median dust temperature of our data points and the red dashed line, in bins of $0.2\,$dex. 
}
\end{figure*}

\begin{figure*}
\center
\includegraphics[width=13.5cm]{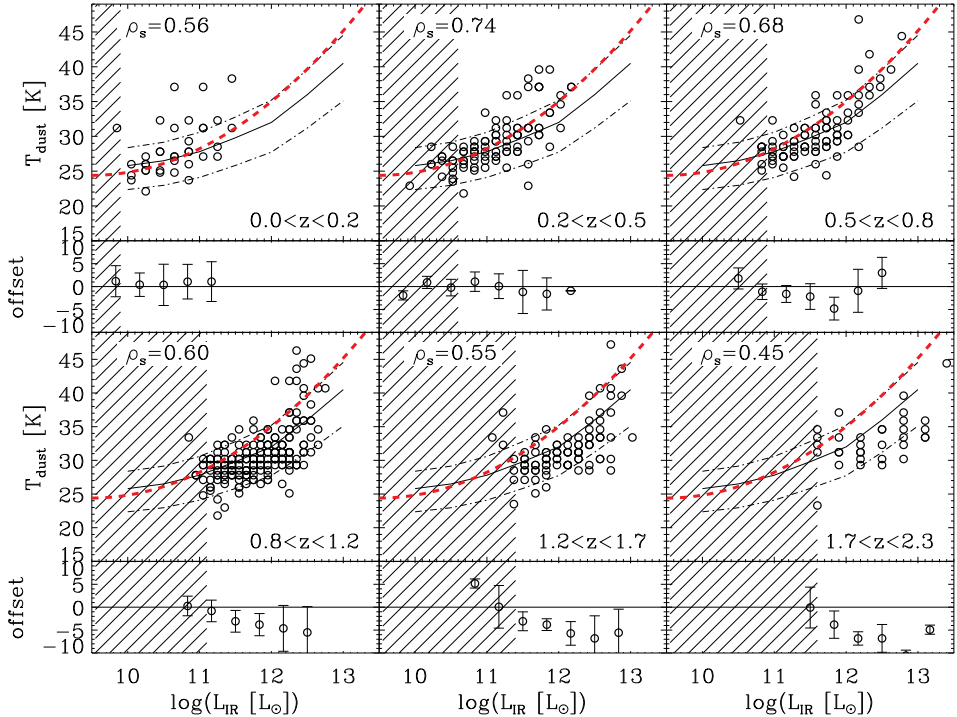}
\caption{ \label{fig:tdust lir}
Dust temperature of galaxies as a function of $L_{{\rm IR}}$, as derived from our stacking analysis.
In each panel, we give the Spearman correlation factor derived from our data points.
Red dashed lines correspond to a second order polynomial fit to the data points of the $0.2<z<0.5$ redshift bin. 
Hatched areas represented the regions of parameter space affected by incompleteness (see text and Fig.~\ref{fig:flag}).
The \citet{chapman_2003} derivation of the median and interquartile range of the relation observed at $z$$\,\thicksim\,$$0$ is shown by solid and dot-dashed lines, linearly extrapolated to $10^{13}\,L_{\odot}$. 
The lower panel of each redshift bin shows the offset between the median dust temperature of our data points and the red dashed line, in bins of $0.3\,$dex. 
}
\end{figure*}

\begin{figure*}
\center
\includegraphics[width=13.5cm]{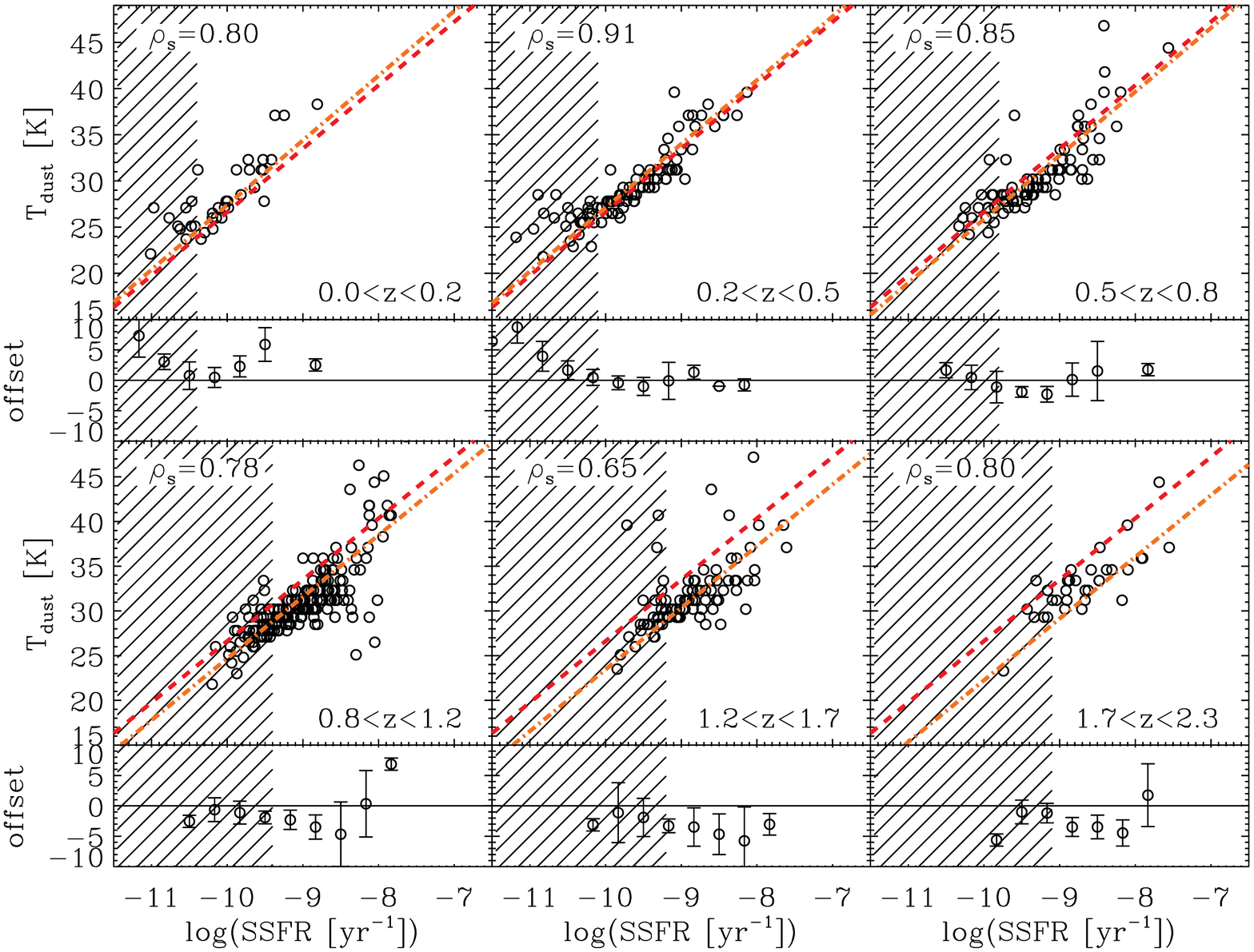}
\caption{ \label{fig:tdust ssfr}
Dust temperature of galaxies as a function of SSFR, as derived from our stacking analysis.
In each panel, we give the Spearman correlation factor derived from our data points.
Red dashed lines correspond to a linear fit to the data points of the $0.2<z<0.5$ redshift bin. 
Orange dot-dashed lines show the redshift evolution of the $T_{\rm dust}$$-$SSFR relation, fixing its slope to that observed in the $0.2<z<0.5$ redshift bin (i.e., slope of the red dashed line) and fitting its zero point with a $A\times(1+z)^{B}$ function (see Eq. \ref{eq:T vs ssfr relation}).
Hatched areas represent the regions of parameter space affected by incompleteness (see text and Fig. \ref{fig:flag}).
The lower panel of each redshift bin shows the offset between the median dust temperature of our data points and the red dashed line, in bins of $0.3\,$dex. 
}
\end{figure*}

To investigate which parameter best correlates with the dust temperature of galaxies, we show in Figs. \ref{fig:tdust dssfr}, \ref{fig:tdust lir} and \ref{fig:tdust ssfr} the variation of the dust temperature of galaxies as function of their $\Delta$log$({\rm SSFR})_{\rm MS}$, $L_{{\rm IR}}$ and ${\rm SSFR}$, respectively.
Because individual detections do not probe large ranges in the $\Delta$log$({\rm SSFR})_{\rm MS}$, $L_{{\rm IR}}$ and ${\rm SSFR}$ parameter spaces, here we only rely on dust temperatures inferred from our stacking analysis. 
In these figures, the dashed regions indicate regimes of parameter space affected by large incompleteness as defined in Fig. \ref{fig:flag}.
Constraints in these regions have to be treated with caution.

In each redshift interval, the dust temperature of galaxies better correlates with their $\Delta$log$({\rm SSFR})_{\rm MS}$ and SSFR values than with $L_{{\rm IR}}$, as revealed by the Spearman correlation factors of the relations between $T_{{\rm dust}}$ and these three quantities.
The Spearman correlation factors of the $T_{{\rm dust}}$$-$$\Delta$log$({\rm SSFR})_{\rm MS}$ and $T_{{\rm dust}}$$-$SSFR correlations are statically equivalent in all our redshift bins.
From this analysis we can conclude that the dust temperature of galaxies is more fundamentally linked to their SSFR and $\Delta$log$({\rm SSFR})_{\rm MS}$ than to their $L_{{\rm IR}}$.

To go further in our understanding of the $T_{{\rm dust}}$$-$$L_{{\rm IR}}$, $T_{{\rm dust}}$$-$$\Delta$log$({\rm SSFR})_{\rm MS}$ and $T_{{\rm dust}}$$-$SSFR correlations, we  also study their redshift evolution.  
We first fit the $T_{{\rm dust}}$$-$$L_{{\rm IR}}$, $T_{{\rm dust}}$$-$$\Delta$log$({\rm SSFR})_{\rm MS}$ and the $T_{{\rm dust}}$$-$SSFR correlations in the $0.2<z<0.5$ redshift bin, where for all three correlations the Spearman correlation factor is highest, and the region of parameter space reliably probed is largest.
The $T_{{\rm dust}}$$-$$L_{{\rm IR}}$ correlation is fitted with a second order polynomial function, while the $T_{{\rm dust}}$$-$$\Delta$log$({\rm SSFR})_{\rm MS}$ and the $T_{{\rm dust}}$$-$SSFR correlations are fitted with linear functions.
We note that \citet{chapman_2003} described the $T_{{\rm dust}}$$-$$L_{{\rm IR}}$ correlation using a double power-law function instead of a second order polynomial.
However, this specific choice does not affect our results.  

Once these best-fitting relations are established in the most reliable redshift interval, the fits are compared with the other intervals to track the redshift evolution of the normalization of these relations.
The $T_{{\rm dust}}$$-$$\Delta$log$({\rm SSFR})_{\rm MS}$ correlation seems to smoothly evolve from $z$$\,=\,$$0$ to $z$$\,=\,$$2.3$, with dust temperatures increasing with redshift at fixed $\Delta$log$({\rm SSFR})_{\rm MS}$. 
Similarly, the $T_{{\rm dust}}$$-$SSFR correlation smoothly evolves up to $z$$\,\thicksim\,$$2$, with the dust temperature of galaxies evolves towards colder values at high redshift at fixed SSFR.
Finally, the $T_{{\rm dust}}$$-$$L_{{\rm IR}}$ correlation slightly evolves from $z$$\,=\,$$0$ to $z$$\,=\,$$2.3$, with high redshift galaxies exhibiting colder dust temperatures than their low redshift counterparts. We note that the $T_{{\rm dust}}$$-$$L_{{\rm IR}}$ correlations found here are consistent with those of \citet{symeonidis_2013} which are based on individual detections.


To reproduce the redshift evolution of the $T_{{\rm dust}}$$-$$\Delta$log$({\rm SSFR})_{\rm MS}$ and $T_{{\rm dust}}$$-$SSFR relations, we fixed their slopes to those measured in the $0.2<z<0.5$ redshift bin and fitted their zero points using a $A\times(1+z)^{B}$ function.
For the $T_{{\rm dust}}$$-$$\Delta$log$({\rm SSFR})_{\rm MS}$ relation, we find
\begin{equation}
\label{eq:T vs dssfr relation}
T_{\rm dust}=26.5\times(1+z)^{0.18}+6.5\times\Delta$log$({\rm SSFR})_{\rm MS},
\end{equation}
while for the $T_{{\rm dust}}$$-$SSFR relation we find
\begin{equation}
\label{eq:T vs ssfr relation}
T_{\rm dust}=98\times(1+z)^{-0.065}+6.9\times{\rm log(SSFR)}
\end{equation}
Because the $T_{{\rm dust}}$$-$${\rm \Delta log(SSFR)_{\rm MS}}$ and $T_{{\rm dust}}$$-$SSFR correlations are statistically very significant, they supersede the $T_{{\rm dust}}$$-$$L_{\rm IR}$ correlation classically used to predict the dust temperature of galaxies.
In particular, Eq. \ref{eq:T vs ssfr relation}, which only relies on the stellar masses, SFRs and redshifts of galaxies, could be used to improve far-infrared predictions from semi-analytical or backward evolutionary models.
\section{Discussion\label{sec:discussion}}

Using deep \textit{Herschel} observations and a careful stacking analysis, we found, over a broad range of redshift, that the dust temperature of galaxies better correlates with their SSFRs or their $\Delta$log$({\rm SSFR})_{\rm MS}$ values than with $L_{\rm IR}$.
This finding supersedes the $T_{{\rm dust}}$$-$$L_{\rm IR}$ correlation classically implemented in far-infrared SED template libraries.
These results also provide us with important information on the conditions prevailing in the star-forming regions of galaxies and the evolution of these conditions as a function of redshift.

\subsection{$T_{\rm dust}$ variations in the SFR$-M_{\ast}$ plane at fixed redshift\label{subsec:tdust at fixed redshift}}

In a given redshift bin, our results unambiguously reveal that the dust temperature of galaxies correlates with their SSFR or equivalently with their $\Delta$log$({\rm SSFR})_{\rm MS}$ (at a given redshift, SSFR and $\Delta$log$({\rm SSFR})_{\rm MS}$ are nearly equivalent because the slope of the MS is close to unity; log(SFR$_{\rm MS})={\rm log(}\,M_{\ast})+C(z)$, so log[SSFR$_{\rm MS}(M_{\ast},z)]=C(z)$ and ${\rm \Delta log(SSFR)_{\rm MS}=log[SSFR(galaxy)]}-C(z)$).
The universality of the dust temperature at fixed $\Delta$log$({\rm SSFR})_{\rm MS}$ indicates that galaxies with a given $\Delta$log$({\rm SSFR})_{\rm MS}$ are composed of the same type of star-forming regions and that the increase of SFR with stellar mass is only due to an increase of the number of such star-forming regions. 
In that picture, galaxies situated on the MS are dominated by star-forming regions with relatively low radiation fields and thus relatively cold dust temperatures, while galaxies situated far above the MS are dominated by star-forming regions exposed to extremely high radiation fields, yielding hotter dust temperatures.

As a first approximation, and in the case where star-forming regions are optically thick in the rest-frame UV and optically thin in the rest-frame far-infrared, one can link the dust temperature of a galaxy to the radiation field seen per unit of dust mass: 
\begin{equation}
\label{eq:tdust}
L_{\rm IR}/M_{\rm dust}\,\propto\,T_{\rm dust}^{4+\beta},
\end{equation}
i.e., the dust temperature of a galaxy increases if the input radiation is delivered to fewer dust grains. 
In that approximation, one can then link the dust temperature of galaxies to the radiation field seen per unit of gas mass (i.e., $L_{\rm IR}/M_{\rm gas}$), via the gas-to-dust ratio relation:
\begin{equation}
\label{eq:gasdust}
{\rm log(}M_{\rm gas}/M_{\rm dust})=-0.85\times\mu+9.4,
\end{equation}
where $\mu$ is the metallicity \citep{leroy_2011}.
At fixed stellar mass (equivalent to fixed metallicity, assuming a mass-metallicity relation), the smooth increase of dust temperature with $\Delta$log$({\rm SSFR})_{\rm MS}$ could thus be interpreted as an increase of $L_{\rm IR}/M_{\rm gas}$, i.e., the star-formation efficiency (SFE) of galaxies.
Using direct molecular gas observations, several studies have effectively observed a significant increase of the SFE with $\Delta$log$({\rm SSFR})_{\rm MS}$ \citep[][]{saintonge_2011a,saintonge_2011b,saintonge_2012,genzel_2010,daddi_2010}.
In their local Universe sample, \citet{saintonge_2012} found
\begin{equation}
\label{eq:saintonge}
{\rm log(SFE)}\,=\,0.5\times\Delta{\rm log(SSFR})_{\rm MS}+C_1,
\end{equation}
 for $-0.5$$\,<\,$$\Delta$log$({\rm SSFR})_{\rm MS}$$\,<\,$$1$.
Combining equations \ref{eq:tdust}, \ref{eq:gasdust} and \ref{eq:saintonge}, assuming that $\Delta{\rm log(SSFR})_{\rm MS}$ and $\mu$ are independent, and assuming that the gas-to-dust relation of \citet{leroy_2011} holds at high redshift, we can thus predict from the gas phase the evolution of the dust temperature as a function of $\Delta$log$({\rm SSFR})_{\rm MS}$ and $\mu$:
\begin{equation}
\label{eq:variation}
{\rm log(}T_{\rm dust})=0.5\times1/(4+\beta)\times\Delta{\rm log(SSFR})_{\rm MS} -0.85\times1/(4+\beta)\times\mu+C_2,
\end{equation}

Over the range of stellar masses probed by our sample (i.e., $\thicksim1.5\,$dex), we expect, based on the mass-metallicity relation, an increase in metallicity of $\thicksim0.25\,$dex ($\thicksim0.4\,$dex) at $z=0$ ($z=2$) (Tremonti et al. \citeyear{tremonti_2004}; $z=2$, Erb et al. \citeyear{erb_2006}; see also Genzel et al. \citeyear{genzel_2012}).
At fixed $\Delta$log$({\rm SSFR})_{\rm MS}$, the dust temperatures of galaxies with high stellar masses should thus be lower by a factor 1.09 (1.15) than those of galaxies with low stellar masses (see Eq. \ref{eq:variation} and assuming $\beta=1.5$).
On the MS, this corresponds to a dust temperature variation of $\thicksim\,$$2\,$K ($5\,$K).
Differentiating galaxies by stellar masses in Fig. \ref{fig:tdust dssfr} supports this prediction.
As a consequence, we can assume that:
\begin{equation}
\label{eq:variation finale}
 {\rm log(}T_{\rm dust})=0.5\times1/(4+\beta)\times\Delta{\rm log(SSFR})_{\rm MS}+C_3, 
 \end{equation}
with variations in metallicities (equivalently $M_{\ast}$) creating a dispersion of a few degrees around this relation.
In Figure~\ref{fig:tdust dssfr}, we compare the prediction from Eq.~\ref{eq:variation finale}, assuming $\beta=1.5$, with the observed $T_{{\rm dust}}$$-$$\Delta$log$({\rm SSFR})_{\rm MS}$ relations (see green triple-dot-dashed lines).
There is a very good agreement between the slope of our predictions and the observations.
We conclude that variations in dust temperature with $\Delta$log$({\rm SSFR})_{\rm MS}$ can be, qualitatively and quantitatively, explained by the variation of the SFE with $\Delta$log$({\rm SSFR})_{\rm MS}$, as observed by \citet{saintonge_2012}.
We note that the smooth evolution of the dust temperature (equivalently SFE) with $\Delta$log$({\rm SSFR})_{\rm MS}$ could correspond to a smooth evolution of the ISM conditions with $\Delta$log$({\rm SSFR})_{\rm MS}$, or to a linear combination of two types of star-forming regions, i.e.: (i) cold star-forming regions formed in secularly evolving galaxies; and (ii) hot star-forming regions formed in starbursting galaxies via large gravitational instabilities (e.g., major mergers).
Unfortunately, using our unresolved observations we cannot discriminate between these two interpretations.

We note that \citet{magdis_2012} report no dust temperature variations between galaxies situated on the lower and upper part of the MS (i.e., $-0.4$$\,<\,$$\Delta$log$({\rm SSFR})_{\rm MS}$$\,<\,$$0.4$).
Based on this finding, they argue that across the MS, the main parameter controlling $\Delta$log$({\rm SSFR})_{\rm MS}$ is $f_{\rm gas}$$\,\equiv\,$$M_{\rm gas}/M_{\ast}$ (i.e., SFE being roughly constant): at fixed stellar mass, galaxies situated on the lower part of the MS would contain less molecular gas (i.e., have lower $f_{\rm gas}$) than galaxies situated on the upper part of the MS.
While \citet{saintonge_2012} qualitatively confirmed the increase of $f_{\rm gas}$ across the MS, they also found a significant increase in SFE which leads to the increase of $T_{{\rm dust}}$ across the MS found in our sample.
Thus, our results, as well as those of \citet{saintonge_2012}, support the hypothesis that the localization of a galaxy across the MS is not only controlled by $f_{\rm gas}$ but rather by both SFE and $f_{\rm gas}$.


\subsection{$T_{\rm dust}$ variations in the SFR$-M_{\ast}$ plane as function of the redshift}

While the slope of the $T_{{\rm dust}}$$-$$\Delta$log$({\rm SSFR})_{\rm MS}$ correlation is the same in all our redshift bins, its normalization evolves towards hotter dust temperatures at higher redshift.
In our simple description, the dust temperature of a galaxy only depends on its SFE and metallicity; therefore one should also observe variations of these parameters with redshift.

In the stellar mass range of our study (i.e., $10^{10}\,$M$_{\odot}$$\,<\,$$M_{\ast}$$\,<\,$$10^{11.5}$), the metallicity $\mu$ of galaxies systematically decreases by $\thicksim\,$$0.5\,$dex from $z=0$ to $z=2$ ($z=0$, Tremonti et al. \citeyear{tremonti_2004}; $z=2$, Erb et al. \citeyear{erb_2006}; Shapley et al. \citeyear{shapley_2005}; Liu et al. \citeyear{liu_2008}; Zahid et al. \citeyear{zahid_2013}; see also Genzel et al. \citeyear{genzel_2012}).
At fixed $\Delta$log$({\rm SSFR})_{\rm MS}$ and assuming $\beta=1.5$, this decrease of the metallicity should translate into an increase of the dust temperature by a factor $\thicksim\,$$1.2$. For MS galaxies, this corresponds to a rise from $26\,$K to $31\,$K between $z=0$ to $z=2$ (see eq. \ref{eq:variation}).
More generally, on the MS we can predict: 
\begin{equation}
\label{eq:Tdust MS}
{\rm log(}T_{\rm dust}^{\rm MS}[z])=-0.85\times1/(4+\beta)\times\Delta\mu[z]+C_4,
\end{equation}
where  $\Delta\mu[z]=\,-1.05\times\,$log(1+$z$) describes the mean increase of the metallicity with redshift over the stellar mass range probed by our sample. 
This formula is used to normalize the predicted $T_{{\rm dust}}$$-$$\Delta$log$({\rm SSFR})_{\rm MS}$ correlation, assuming $T_{\rm dust}^{\rm MS}[z=0]=26\,$K and $\beta=1.5$ (see the green triple-dot-dashed lines in Fig. \ref{fig:tdust dssfr}).
Our predictions are perfectly in line with the observations.
We thus conclude that the evolution of metallicity with redshift might be the main driver of the normalization of the $T_{{\rm dust}}$$-$$\Delta$log$({\rm SSFR})_{\rm MS}$ correlation.
We note that in the local Universe an evolution of the dust temperature with metallicity has already been observed \citep{engelbracht_2008}.
In that study the slope of the $T_{{\rm dust}}$$-$metallicity relation roughly follows our predictions, i.e., log($T_{{\rm dust}}$)$\,=\,$$-0.85\times1/(4+\beta)\times\mu+C_4$.

\citet{tacconi_2013} found that the SFE of MS galaxies evolves from $z=0$ to $z=1.5$ following:
\begin{equation}
\label{eq:Tdust MS SFE}
{\rm log(SFE_{MS})}={\rm log}(1+z)+C_{5},
\end{equation}
This increase of the SFE should translate into an increase of the dust temperature by a factor $\thicksim\,$$1.2$ between $z=0$ to $z=2$, i.e., for MS galaxies rising from $26\,$K to $31\,$K.
More generally, on the MS we can predict: 
\begin{equation}
\label{eq:Tdust MS}
{\rm log(}T_{\rm dust}^{\rm MS}[z])=1/(4+\beta)\times{\rm log}(1+z)+C_6,
\end{equation}
This formula is also used to normalize the predicted $T_{{\rm dust}}$$-$$\Delta$log$({\rm SSFR})_{\rm MS}$ correlation, assuming $T_{\rm dust}^{\rm MS}[z=0]=26\,$K and $\beta=1.5$ (i.e., light blue long-dashed lines in Fig. \ref{fig:tdust dssfr}).
Because these predictions are also in line with our observations, we conclude that the evolution of the SFE with redshift might also be the main driver of the normalization of the $T_{{\rm dust}}$$-$$\Delta$log$({\rm SSFR})_{\rm MS}$ correlation.

Combining the evolution of both the SFE and metallicity with redshift, we can predict,
\begin{equation}
\label{eq:Tdust MS tot}
{\rm log(}T_{\rm dust}^{\rm MS}[z])=(0.85\times1.05+1)/(4+\beta)\times{\rm log}(1+z)+C_7,
\end{equation}
In Figure \ref{fig:tdust dssfr}, the dark dot-dashed lines present the predictions from this formula, assuming $T_{\rm dust}^{\rm MS}[z=0]=26\,$K and $\beta=1.5$.
We observe that it overestimates the normalization with redshift of the $T_{{\rm dust}}$$-$$\Delta$log$({\rm SSFR})_{\rm MS}$ correlation.\\

In a simple optically thin model where $T_{\rm dust}$ directly traces the galaxy-integrated $L_{\rm IR}$ and $M_{\rm dust}$, our findings can be summarized as follows:\\
$-$ At fixed redshift, the main parameter controlling the variation of $T_{\rm dust}$ with $\Delta$log$({\rm SSFR})_{\rm MS}$ is the SFE of galaxies: galaxies with high $\Delta$log$({\rm SSFR})_{\rm MS}$ are more efficient in turning their gas into stars, which leads to higher input power per unit of dust mass and therefore higher dust temperatures.\\
$-$ At fixed $\Delta$log$({\rm SSFR})_{\rm MS}$, the increase of the metallicity with the stellar masses can explain some of the scatter observed in the $T_{{\rm dust}}$$-$$\Delta$log$({\rm SSFR})_{\rm MS}$ and $T_{{\rm dust}}$$-$${\rm SSFR}$ correlations.\\
$-$ The normalization with redshift of the $T_{{\rm dust}}$$-$${\rm \Delta log(SSFR)_{\rm MS}}$ correlation towards hotter dust temperatures can be explained by the global decrease of the metallicity of galaxies with redshift or by the global increase of the SFE of galaxies with redshift. However, combining the evolution of both the metallicity and SFE with redshift, we overestimate the normalization towards hotter dust temperatures of the $T_{\rm dust}$-$\Delta$log$({\rm SSFR})_{\rm MS}$ correlation.
This may indicate that the assumptions of our simple model do not hold. 
In particular, the scaling relations that we adopted for the evolution of the metallicity and SFE with redshift and with offset from the MS might be overall inconsistent. 
This will need future revision.\\

We note that using a sample of lensed galaxies to extend the study of the dust temperatures of MS galaxies to lower stellar masses and higher redshifts ($4\times10^9<$M$_{\ast}<10^{11}\,$M$_{\odot}$,  $2<z<3$), \citet{saintonge_2013} also report higher $T_{\rm dust}$ values compared to lower redshift samples, and attribute this variation to lower metallicities and increased star formation efficiencies in high redshift, low mass galaxies.

Finally, we note that \citet{nordon_2012} and \citet{elbaz_2011} found the PAH$-$to$-$$L_{\rm IR}$ ratio, parametrized as $L_{8}/L_{\rm IR}$, to strongly correlate with $\Delta$log$({\rm SSFR})_{\rm MS}$.  This correlation appears to remain the same up to $z\thicksim2$.
Therefore, PAHs and dust temperatures do not evolve in the same manner as a function of redshift.
This situation could arise from differences in the mechanics heating the PAHs and the dust components dominating the FIR emission of galaxies.
In the model of \citet{draine_2007}, the fraction of power radiated into the $7.9\,\mu$m band by the single-photon heated PAHs, observed as $L_{8}/L_{\rm IR}$, is essentially proportional to $q_{\rm PAH}$, the fraction of dust mass in the form of PAH grains with less than $10^3$ carbon atoms.  It does not appear to scale with $\langle U\rangle$, the mean intensity seen per unit of dust mass \citep[see figures 19c and 19d of][]{draine_2007}.
Consequently, in this model, a strong increase of the dust temperature can occur while the PAH$-$to$-$$L_{\rm IR}$ ratio remains constant, if $q_{\rm PAH}$ remains the same and the intensity of the radiation field seen by the dust increases.
The question is thus: at fixed $\Delta$log$({\rm SSFR})_{\rm MS}$, \textit{is $q_{\rm PAH}$ affected by the evolution of the metallicity with redshift or by the evolution of the SFE with redshift$\,$?}
In local star-forming galaxies and in the metallicity range of our sample ($\mu>8.1$), there is only a weak correlation between the metallicity and the PAH$-$to$-$$L_{\rm IR}$ ratio \citep{smith_2007,engelbracht_2008,hunt_2010}.
Thus, the evolution of the metallicity with redshift might affect the normalization of the $T_{{\rm dust}}$$-$$\Delta$log$({\rm SSFR})_{\rm MS}$ correlation but not significantly affect $q_{\rm PAH}$ and thus leads to the universal (PAH$-$to$-$$L_{\rm IR}$)-$\Delta$log$({\rm SSFR})_{\rm MS}$ correlation observed in \citet{elbaz_2011} and \citet{nordon_2012}. 
For the SFE, the situation is somewhat different. 
At fixed redshift, we have explained the $T_{{\rm dust}}$$-$$\Delta$log$({\rm SSFR})_{\rm MS}$ correlation using the SFE$-$$\Delta$log$({\rm SSFR})_{\rm MS}$ correlation observed by \citet{saintonge_2012}.
Following a similar train of thought, one could thus also ascribe the (PAH$-$to$-$$L_{\rm IR}$)-$\Delta$log$({\rm SSFR})_{\rm MS}$ correlation to the SFE$-$$\Delta$log$({\rm SSFR})_{\rm MS}$ correlation.
In this case, however, one would expect to observe an evolution with redshift of the normalization of both the $T_{{\rm dust}}$$-$$\Delta$log$({\rm SSFR})_{\rm MS}$ and (PAH$-$to$-$$L_{\rm IR}$)-$\Delta$log$({\rm SSFR})_{\rm MS}$ correlations, following the evolution with redshift of the normalization of the SFE$-$$\Delta$log$({\rm SSFR})_{\rm MS}$ correlation.
Is the universality of the (PAH$-$to$-$$L_{\rm IR}$)-$\Delta$log$({\rm SSFR})_{\rm MS}$ correlation to be interpreted as a proof that actually the SFE$-$$\Delta$log$({\rm SSFR})_{\rm MS}$ correlation does not significantly evolve with redshift$\,$? And that the evolution of the metallicity with redshift is the main driver for the normalization of the $T_{{\rm dust}}$$-$$\Delta$log$({\rm SSFR})_{\rm MS}$ correlation$\,$?
To obtain answers to these questions, additional studies are required.\\

\subsection{Towards a better understanding of the MS of star-formation}

The existence of the MS of star formation is currently interpreted as evidence that the bulk of the galaxy population is forming stars gradually with long duty cycles.
In this picture, MS galaxies evolve with a secular mode of star formation, likely sustained by a continuous gas accretion from the IGM and along the cosmic web \citep{dekel_2009,dave_2010}, while star-forming galaxies located far above the MS evolve through strong starbursts with short duty-cycles, mainly triggered by major mergers.
The strong correlation observed between $T_{{\rm dust}}$ and $\Delta$log$({\rm SSFR})_{\rm MS}$ gives weight to this interpretation.
It unambiguously reveals that the physical conditions prevailing in the star-forming regions of on- and far-above-MS galaxies are not the same.
Furthermore, because the $T_{{\rm dust}}$$-$$\Delta$log$({\rm SSFR})_{\rm MS}$ correlation can be understood as a variation of the SFE with $\Delta$log$({\rm SSFR})_{\rm MS}$, it also supports the hypothesis of different modes of star-formation for these two populations of galaxies.
Unfortunately, using our unresolved observations we cannot distinguish if galaxies lying at intermediate distance from the MS correspond to a linear combination or to a smooth evolution of the ISM conditions between these two extreme cases.

In line with our findings, \citet{magnelli_2012b} found that on- and above-MS galaxies exhibit different CO-to-H$_{2}$ conversion factors (i.e., $\alpha_{{\rm CO}}$), indicating differences in the physical conditions prevailing in their Giant Molecular Clouds \citep[i.e. GMCs,][]{genzel_2010,daddi_2010}.
MS galaxies have high $\alpha_{{\rm CO}}$ factors, consistent with star-forming regions mainly composed of well virialized GMCs, while above-MS galaxies have low $\alpha_{{\rm CO}}$ factors consistent with star-forming regions being mainly un-virialized GMCs as observed in local major-mergers.
Combined with other independent observational properties, the current vision of the main-sequence of star-formation can be summarized as follows.
Galaxies situated on the MS of star-formation have disk-like morphology \citep{wuyts_2011b}, have star-forming regions mainly composed of virialized GMCs \citep{magnelli_2012b}, have relatively low SFR surface density \citep{wuyts_2011b}, high PAH$-$to$-$$L_{\rm IR}$ ratio \citep[][]{elbaz_2011,nordon_2012}, high \ion{C}{II}$-$to$-$$L_{\rm IR}$ ratio \citep{gracia_carpio_2011} and cold dust temperature.
Galaxies situated far above the MS have bulge-like morphology, star-forming regions dominated by un-virialized GMCs, have relatively high SFR surface density, low PAH$-$to$-$$L_{\rm IR}$ ratio, low \ion{C}{II}$-$to$-$$L_{\rm IR}$ and hot dust temperature.
All these independent observations give strong weight to the interpretation where on- and above-MS galaxies evolve through different modes of star-formation, a secular and a starbursting mode, respectively.

\section{Summary\label{sec:conclusion}}
Using deep \textit{Herschel} PACS and SPIRE observations of the GOODS-N, GOODS-S and COSMOS fields, we study variations of the dust temperature of galaxies in the SFR$-M_{\ast}$ plane up to $z\thicksim2$.
Dust temperatures are inferred using the stacked PACS/SPIRE far-infrared photometries of each SFR$-$$M_{\ast}$$-$$z$ bin.
Based on this careful analysis, we constrain the dust temperature of galaxies over a broad range of redshifts, and study the evolution of the $T_{{\rm dust}}$$-$$L_{{\rm IR}}$, $T_{{\rm dust}}$$-$SSFR and $T_{{\rm dust}}$$-$$\Delta$log$({\rm SSFR})_{\rm MS}$ correlations.
Then, using a simple model linking the dust temperature with the dust and gas contents of the ISM, we discuss these correlations in terms of variations of the physical conditions prevailing in the star-forming regions of galaxies.
Our main conclusions are:
\begin{enumerate}
\item Over a broad range of redshifts, the dust temperature of galaxies smoothly increases with their infrared luminosities, SSFRs and ${\rm \Delta log(SSFR)_{\rm MS}}$ values.
The $T_{{\rm dust}}$$-$${\rm SSFR}$ and $T_{{\rm dust}}$$-$${\rm \Delta log(SSFR)_{\rm MS}}$ correlations are statistically much more significant than the $T_{{\rm dust}}$$-$$L_{\rm IR}$ correlation.
The slope of the $T_{{\rm dust}}$$-$${\rm \Delta log(SSFR)_{\rm MS}}$ correlation does not evolve up to $z$$\,\thicksim\,$$2$, but its normalization does: at a given $\Delta$log$({\rm SSFR})_{\rm MS}$, high-redshift galaxies exhibit hotter dust temperatures than their local counterparts.
Similarly, the normalization of the $T_{{\rm dust}}$$-$SSFR correlation evolves towards colder dust temperatures up to $z$$\,\thicksim\,$$2$.
The $T_{{\rm dust}}$$-$$L_{{\rm IR}}$ correlation also evolves with redshift: at fixed $L_{\rm IR}$, high-redshift galaxies exhibit slightly colder dust temperatures than their local counterparts.
\item Because the $T_{{\rm dust}}$$-$${\rm \Delta log(SSFR)_{\rm MS}}$ and $T_{{\rm dust}}$$-$SSFR correlations are statistically very significant, they supersede the $T_{{\rm dust}}$$-$$L_{\rm IR}$ correlation classically used to predict the dust temperatures of galaxies.
These correlations could be used to improve results of semi-analytical or backward evolutionary models.
\item In a simple optically thin model, where $T_{\rm dust}$ directly traces the galaxy-integrated $L_{\rm IR}$ and $M_{\rm dust}$, the slope of the $T_{{\rm dust}}$$-$${\rm \Delta log(SSFR)_{\rm MS}}$ (equivalently SSFR at fixed $z$) correlation can be explained, qualitatively and quantitatively, by the increase of the SFE with $\Delta$log$({\rm SSFR})_{\rm MS}$ found locally by molecular gas studies \citep{saintonge_2012}.
In addition, the normalization with redshift of the $T_{{\rm dust}}$$-$${\rm \Delta log(SSFR)_{\rm MS}}$ correlation can be explained by the global decrease of the metallicity of galaxies with redshift \citep{tremonti_2004,erb_2006} or by the increase of the SFE of MS galaxies with redshift \citep{tacconi_2013}. However, combining the evolution of both the metallicity and SFE with redshift, we overestimate the normalization towards hotter dust temperatures of the $T_{\rm dust}$-$\Delta$log$({\rm SSFR})_{\rm MS}$ correlation.
This may indicate that the assumptions of our simple model do not hold. In particular, the scaling relations that we adopted for the evolution of the metallicity and SFE with redshift and with offset from the MS might be overall inconsistent. This will need future revision.
\end{enumerate}
All these results support the hypothesis that the conditions prevailing in the star-forming regions of MS and far-from-MS galaxies are different.
MS galaxies have star-forming regions with low SFEs and thus cold dust temperatures, while galaxies situated far above the MS seems to be in a starbursting phase characterized by star-forming regions with high SFEs and thus hot dust temperatures.
Galaxies lying at intermediate distance from the MS could correspond, for our unresolved observations, to a linear combination or to a smooth evolution of the physical conditions between these two extreme star-forming regions.
In that picture, at fixed $\Delta$log$({\rm SSFR})_{\rm MS}$, galaxies are dominated by the same type of star-forming regions and the increase of SFR with $M_{\ast}$ is due to an increase of the number of such star-forming regions.

\begin{acknowledgements}
PACS has been developed by a consortium of institutes led by MPE (Germany) and including UVIE (Austria); KU Leuven, CSL, IMEC (Belgium); CEA, LAM (France); MPIA (Germany); INAF-IFSI/OAA/OAP/OAT, LENS, SISSA (Italy); IAC (Spain). This development has been supported by the funding agencies BMVIT (Austria), ESA-PRODEX (Belgium), CEA/CNES (France), DLR (Germany), ASI/INAF (Italy), and CICYT/MCYT (Spain).
SPIRE has been developed by a consortium of institutes led by Cardiff University (UK) and including University of Lethbridge (Canada), NAOC (China), CEA, LAM (France), IFSI, University of Padua (Italy), IAC (Spain), Stockholm Observatory (Sweden), Imperial College London, RAL, UCL-MSSL, UKATC, University of Sussex (UK), Caltech, JPL, NHSC, University of Colorado (USA). This development has been supported by national funding agencies: CSA (Canada); NAOC (China); CEA, CNES, CNRS (France); ASI (Italy); MCINN (Spain); SNSB (Sweden); STFC, UKSA (UK); and NASA (USA).
D.E. received support of the European Community Framework Programme 7, FP7-SPACE Astrodeep, grant agreement no.312725
\end{acknowledgements}
\bibliographystyle{aa}

\begin{appendix}
\section{Far-infrared SEDs}
\begin{figure*}
\center
\includegraphics[width=15cm]{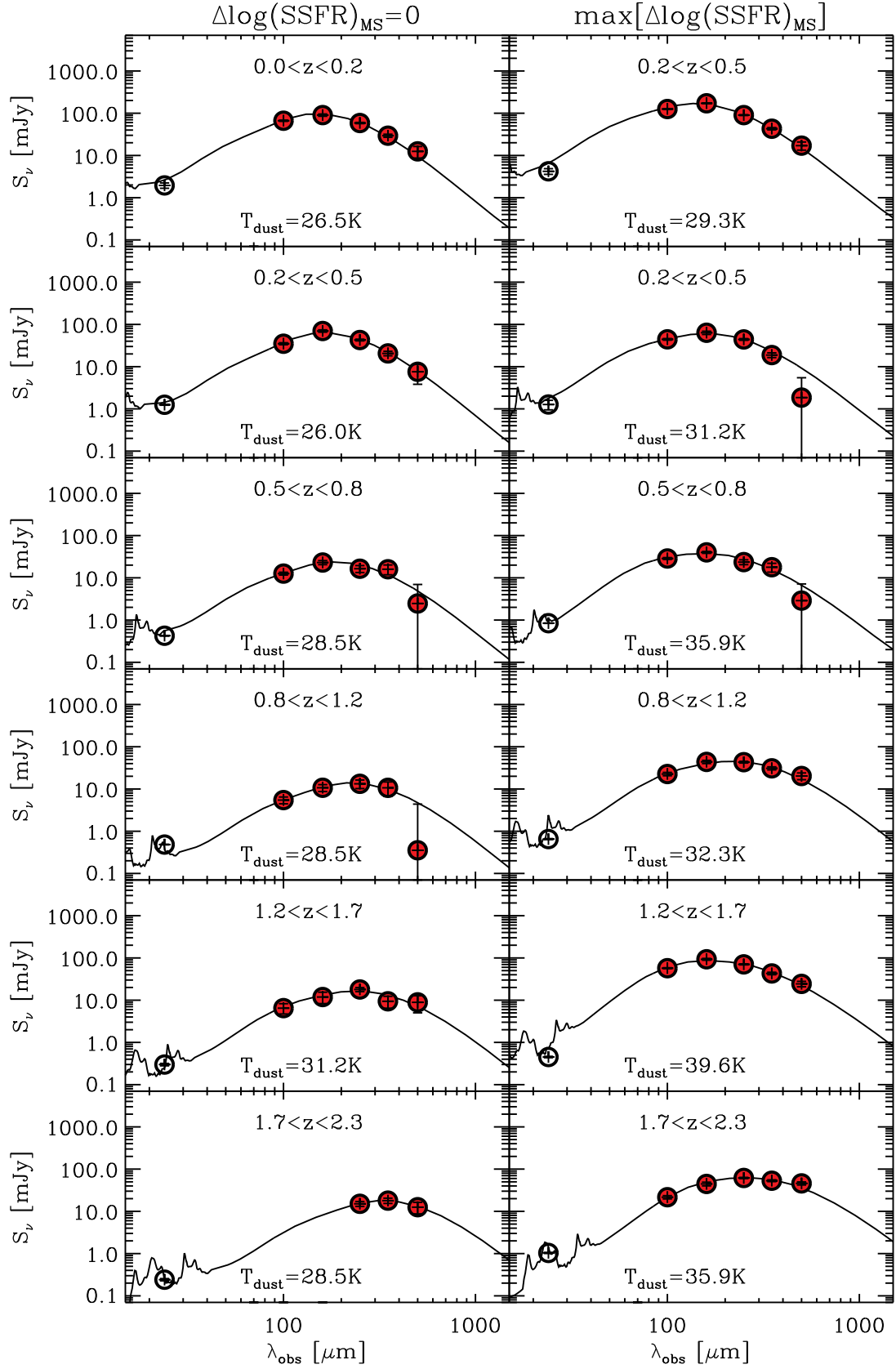}
\caption{ \label{fig:sed detection}
Far-infrared properties of some of our galaxies with individual far-infrared detections.
All these galaxies have $10^{10.8}$$\,<\,$$M_{\ast}\,[{\rm M_{\odot}}]$$\,<\,$$10^{11}$.
In the first column, galaxies have $-0.1$$\,<\,$$\Delta$log$({\rm SSFR})_{\rm MS}$$\,<\,$$0.1$ while in the second column they have $\Delta$log$({\rm SSFR})_{\rm MS}$$\,>\,$${\rm MAX[}\Delta$log$({\rm SSFR})_{\rm MS}]-0.1$.
Black lines show the DH SED templates best-fitting these far-infrared flux densities.
From these fits, we inferred the infrared luminosities and dust temperatures of our galaxies.
}
\end{figure*}

Figure \ref{fig:sed detection} presents the far-infrared properties of some of our galaxies with individual far-infrared detections.
In the first column, galaxies were randomly selected within a sample with $10^{10.8}$$\,<\,$$M_{\ast}\,[{\rm M_{\odot}}]$$\,<\,$$10^{11}$ and $-0.1$$\,<\,$$\Delta$log$({\rm SSFR})_{\rm MS}$$\,<\,$$0.1$.
In the second column, galaxies were randomly selected within a sample with $10^{10.8}$$\,<\,$$M_{\ast}\,[{\rm M_{\odot}}]$$\,<\,$$10^{11}$ and $\Delta$log$({\rm SSFR})_{\rm MS}$$\,>\,$${\rm MAX[}\Delta$log$({\rm SSFR})_{\rm MS}]-0.1$.
To derive the infrared luminosities and dust temperatures of these galaxies (see Sect. \ref{subsec:SFR}, Sect. \ref{subsec:temperature} and Table \ref{tab:DH}), we fitted their far-infrared flux densities with the DH SED template library.
Results of these fits are shown by black lines.

\begin{figure*}
\center
\includegraphics[width=15cm]{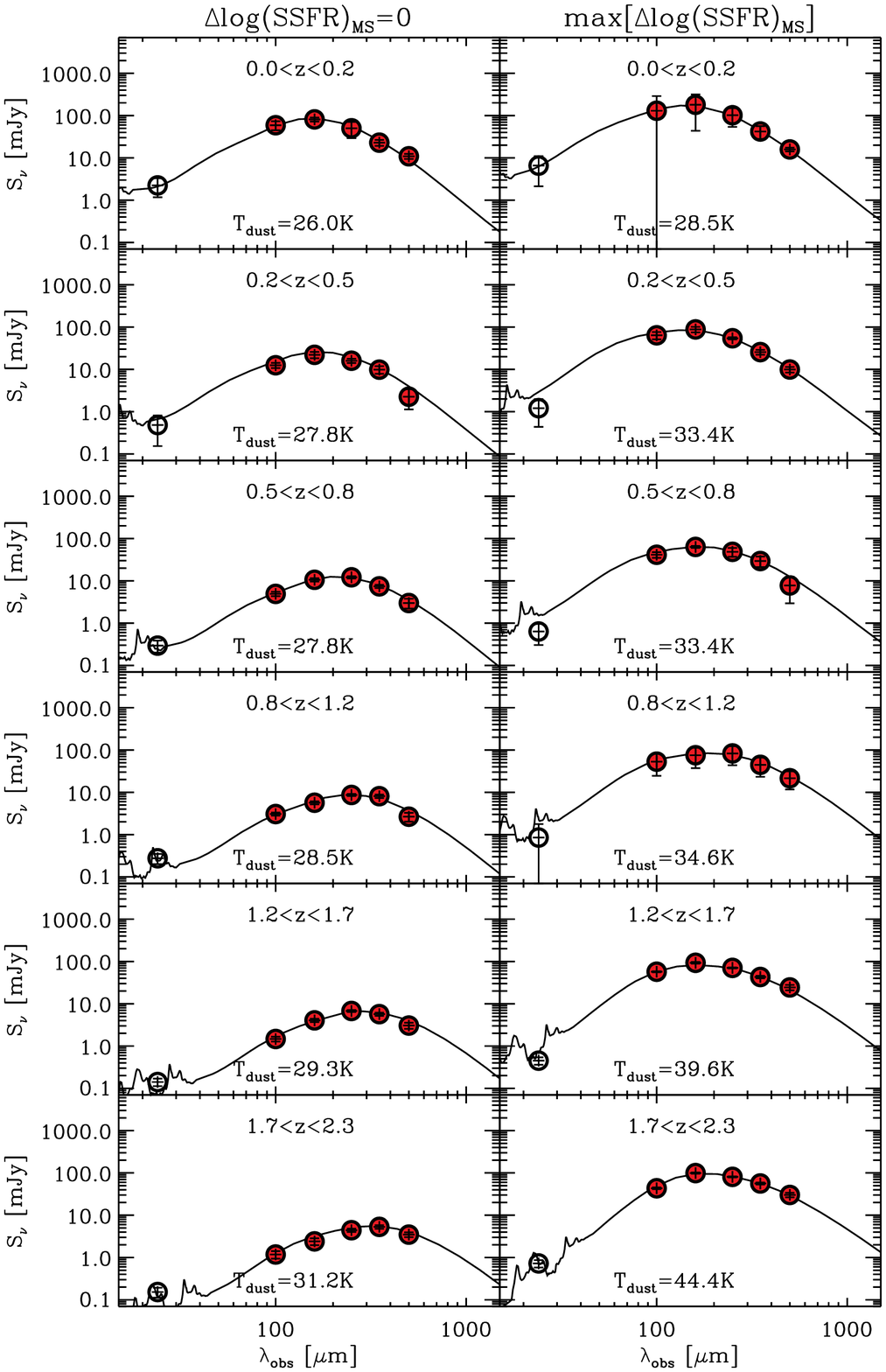}
\caption{ \label{fig:sed stack}
Far-infrared properties of some of our SFR-$M_{\ast}$ bins, as inferred from our stacking analysis.
These SFR-$M_{\ast}$ bins have $10^{10.8}$$\,<\,$$M_{\ast}\,[{\rm M_{\odot}}]$$\,<\,$$10^{11}$.
In the first column, SFR-$M_{\ast}$ bins have $-0.1$$\,<\,$$\Delta$log$({\rm SSFR})_{\rm MS}$$\,<\,$$0.1$, while in the second column they have $\Delta$log$({\rm SSFR})_{\rm MS}$$\,=\,$${\rm MAX[}\Delta$log$({\rm SSFR})_{\rm MS}]$.
Black lines show the DH SED templates best-fitting these far-infrared flux densities.
From these fits, we inferred the mean dust temperatures of galaxies in these SFR-$M_{\ast}$ bins.
We also verified that the infrared luminosities inferred from these fits agree, within 0.3 dex, with the mean infrared luminosities inferred from our ``ladder of SFR indicators'' (see Sect. \ref{subsec:SFR}).
}
\end{figure*}

Figure \ref{fig:sed stack} presents the mean far-infrared properties of some of our SFR-$M_{\ast}$ bins, as inferred from our stacking analysis.
In the first column, SFR-$M_{\ast}$ bins were selected to have $10^{10.8}$$\,<\,$$M_{\ast}\,[{\rm M_{\odot}}]$$\,<\,$$10^{11}$ and $-0.1$$\,<\,$$\Delta$log$({\rm SSFR})_{\rm MS}$$\,<\,$$0.1$, while in the second column, SFR-$M_{\ast}$ bins were selected to have $10^{10.8}$$\,<\,$$M_{\ast}\,[{\rm M_{\odot}}]$$\,<\,$$10^{11}$ and $\Delta$log$({\rm SSFR})_{\rm MS}$$\,=\,$${\rm MAX[}\Delta$log$({\rm SSFR})_{\rm MS}]$.
To derive the mean dust temperature of these SFR-$M_{\ast}$ bins, we fitted their far-infrared flux densities with the DH SED template library (see black lines in Fig. \ref{fig:sed stack}).
\begin{table*}
\caption{\label{tab:DH} Dust temperatures assigned to the DH SED templates by fitting their simulated $z=0$ PACS/SPIRE flux densities with a single modified (i.e., $\beta=1.5$) blackbody function.}
\centering
\begin{tabular}{ c cccc cccc} 
\hline
       $\alpha^{\rm a}$ & 0.062 & 0.125 & 0.188 & 0.250 & 0.312 & 0.375 & 0.438 & 0.500 \\
 $T_{\rm dust}\,$[K] &  49.5 &  49.4 &  49.3 &  49.1 &  49.0 &  48.8 &  48.6 &  48.4 \\
\hline
\\
\hline
       $\alpha^{\rm a}$ & 0.562 & 0.625 & 0.688 & 0.750 & 0.812 & 0.875 & 0.938 & 1.000 \\
 $T_{\rm dust}\,$[K] &  48.2 &  47.9 &  47.6 &  47.2 &  46.8 &  46.3 &  45.8 &  45.1 \\
\hline
\\
\hline
       $\alpha^{\rm a}$ & 1.062 & 1.125 & 1.188 & 1.250 & 1.312 & 1.375 & 1.438 & 1.500 \\
 $T_{\rm dust}\,$[K] &  44.4 &  43.6 &  42.7 &  41.8 &  40.7 &  39.6 &  38.3 &  37.1 \\
\hline
\\
\hline
       $\alpha^{\rm a}$ & 1.562 & 1.625 & 1.688 & 1.750 & 1.812 & 1.875 & 1.938 & 2.000 \\
 $T_{\rm dust}\,$[K] &  35.9 &  34.6 &  33.4 &  32.3 &  31.2 &  30.2 &  29.3 &  28.5 \\
\hline
\\
\hline
       $\alpha^{\rm a}$ & 2.062 & 2.125 & 2.188 & 2.250 & 2.312 & 2.375 & 2.438 & 2.500 \\
 $T_{\rm dust}\,$[K] &  27.8 &  27.1 &  26.5 &  26.0 &  25.5 &  25.1 &  24.8 &  24.4 \\
\hline
\\
\hline
       $\alpha^{\rm a}$ & 2.562 & 2.625 & 2.688 & 2.750 & 2.812 & 2.875 & 2.938 & 3.000 \\
 $T_{\rm dust}\,$[K] &  24.2 &  23.9 &  23.7 &  23.5 &  23.3 &  23.1 &  23.0 &  22.9 \\
\hline
\\
\hline
       $\alpha^{\rm a}$ & 3.062 & 3.125 & 3.188 & 3.250 & 3.312 & 3.375 & 3.438 & 3.500 \\
 $T_{\rm dust}\,$[K] &  22.7 &  22.6 &  22.5 &  22.4 &  22.3 &  22.3 &  22.2 &  22.1 \\
\hline
\\
\hline
       $\alpha^{\rm a}$ & 3.562 & 3.625 & 3.688 & 3.750 & 3.812 & 3.875 & 3.938 & 4.000 \\
 $T_{\rm dust}\,$[K] &  22.1 &  22.0 &  22.0 &  21.9 &  21.9 &  21.8 &  21.7 &  21.6 \\
\hline
\\
\end{tabular}
\begin{list}{}{}
\item[\textbf{Notes.} ]$^{\rm a}$ $\alpha$ is the single parameter of the DH SED template library \citep{dale_2002}. 
\end{list}
\end{table*}


\section{The $T_{{\rm dust}}$$-$$\Delta$log$({\rm SSFR})_{\rm MS}$ relation in the GOODS fields\label{appendix:GOODS}}
\begin{figure*}
\center
\includegraphics[width=13.5cm]{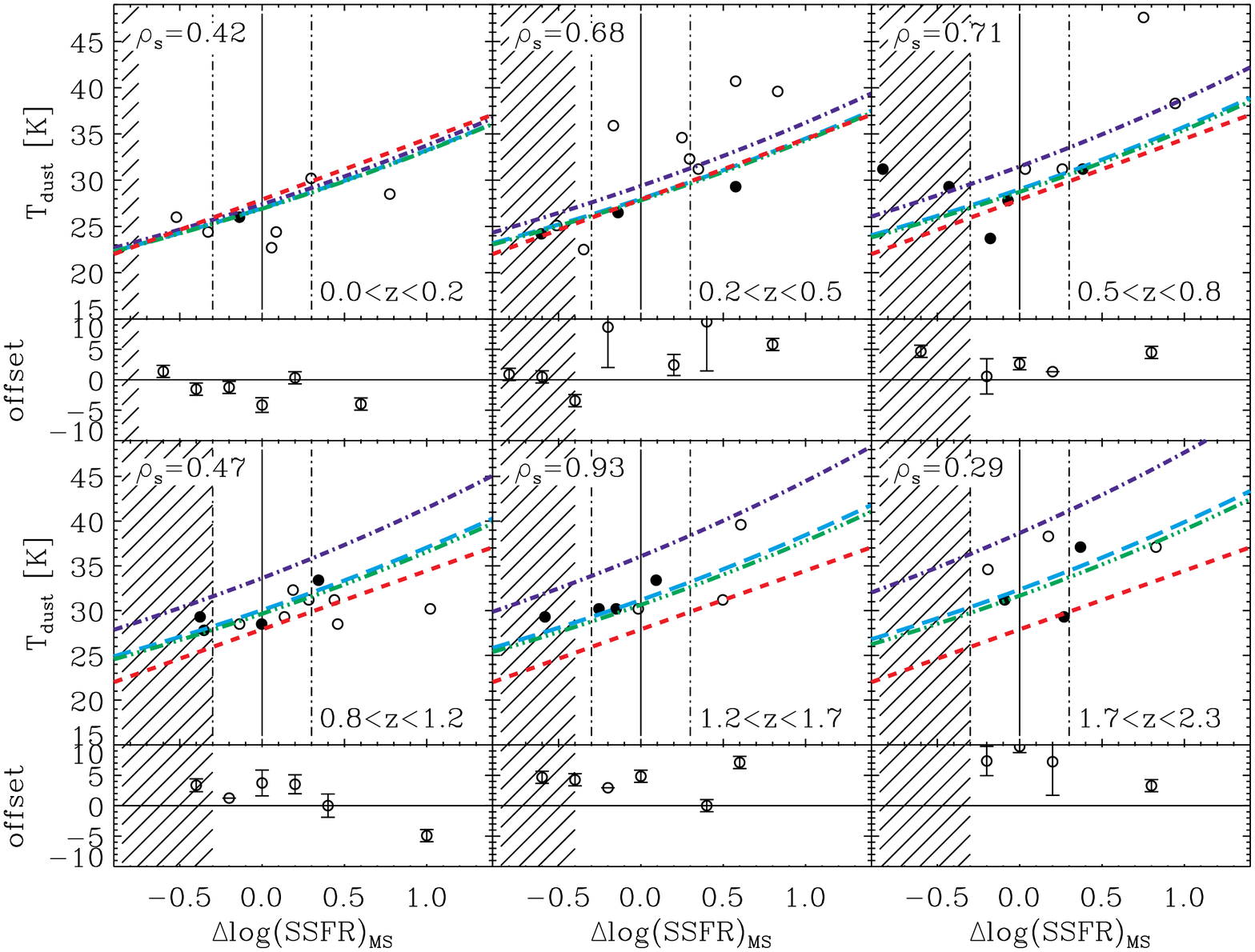}
\caption{ \label{fig:GOODS}
Dust temperature of galaxies as a function of $\Delta$log$({\rm SSFR})_{\rm MS}$, as derived from our stacking analysis using only the GOODS fields (i.e., excluding the COSMOS observations).
Symbols and lines are the same than in Fig. \ref{fig:tdust dssfr}.
}
\end{figure*}

Figure \ref{fig:GOODS} presents the $T_{{\rm dust}}$$-$$\Delta$log$({\rm SSFR})_{\rm MS}$ relation, as inferred using the GOODS fields.
Because we exclude the COSMOS observations from our stacking analysis, our ability to probe and sample properly the SFR-$M_{\ast}$ parameter space decreases.
Therefore, the number of data points available to study the $T_{{\rm dust}}$$-$$\Delta$log$({\rm SSFR})_{\rm MS}$ relation is significantly reduced.
Despite this limitation, we still observe a $T_{{\rm dust}}$$-$$\Delta$log$({\rm SSFR})_{\rm MS}$ relation evolving with redshift.

\section{The $T_{{\rm dust}}$$-$$\Delta$log$({\rm SSFR})_{\rm MS}$ relation using different definition of the MS\label{appendix:MS}}
\begin{figure*}
\center
\includegraphics[width=13.5cm]{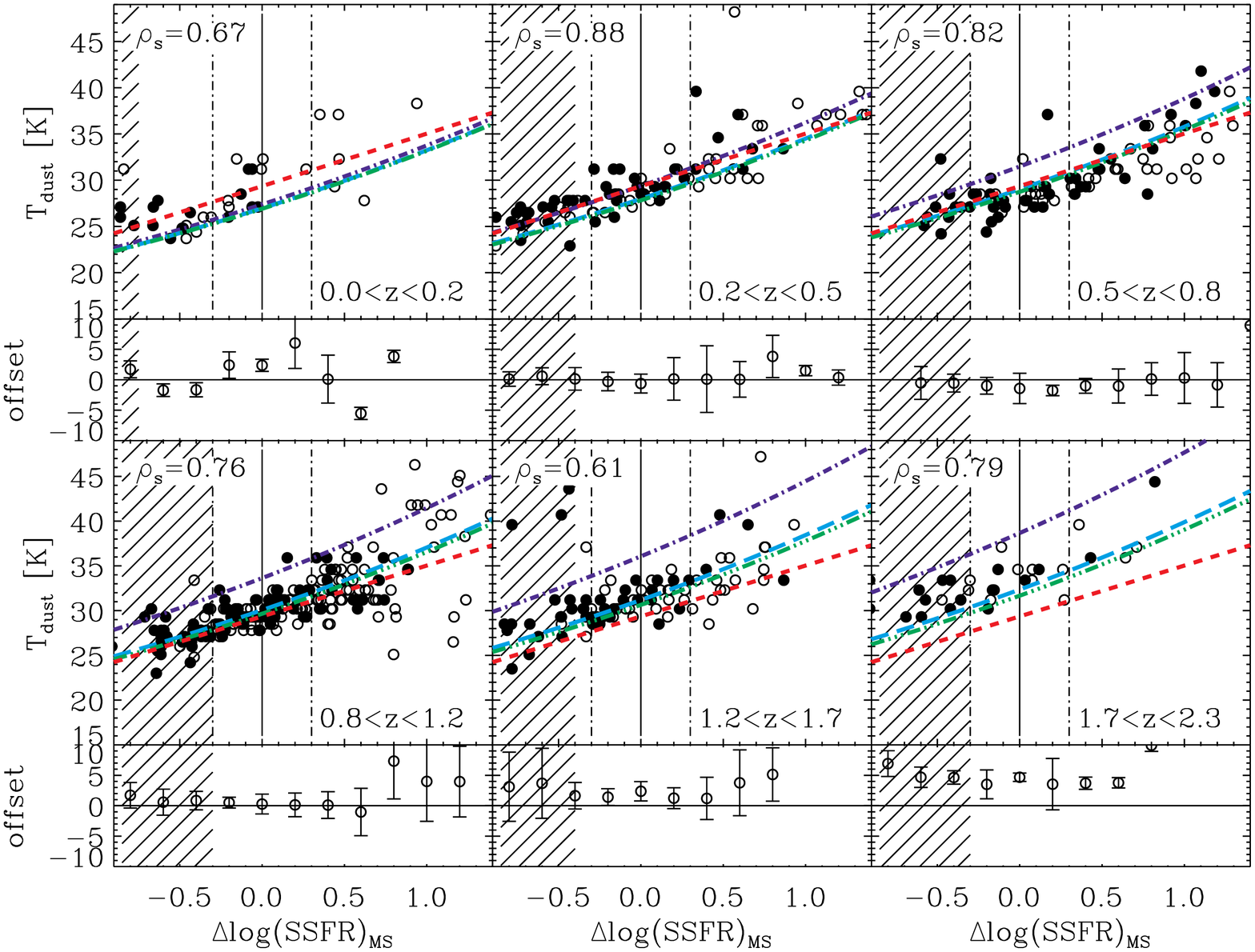}
\caption{ \label{fig:tdust dssfr david}
Dust temperature of galaxies as a function of $\Delta$log$({\rm SSFR})_{\rm MS}$, as derived from our stacking analysis.
In this figure, the definition of the MS of star-formation is taken from \citet{elbaz_2011}.
Symbols and lines are the same than in Fig. \ref{fig:tdust dssfr}.
}
\end{figure*}

In this appendix we test the robustness of the $T_{{\rm dust}}$$-$$\Delta$log$({\rm SSFR})_{\rm MS}$ correlation against changes in the definition of the MS of star-formation.
Figure \ref{fig:tdust dssfr david} shows the $T_{{\rm dust}}$$-$$\Delta$log$({\rm SSFR})_{\rm MS}$ correlation using the definition of the MS from \citet{elbaz_2011}.
We still observe a strong correlation between $T_{{\rm dust}}$ and $\Delta$log$({\rm SSFR})_{\rm MS}$ as revealed by the high Spearman correlation factor found in all our redshift bins.
Our results are thus robust against changes in the definition of the MS.
\section{Authors' affiliations}\label{sect:affiliations}

\begin{enumerate}[label=$^{\arabic{*}}$]
\setcounter{enumi}{9}
\item
Laboratoire AIM, CEA/DSM-CNRS-Universit{\'e} Paris Diderot, IRFU/Service
d'Astrophysique,
B\^at.709, CEA-Saclay, 91191 Gif-sur-Yvette Cedex, France.
\item
California Institute of Technology, 1200 E. California Blvd., Pasadena, CA 91125, USA
\item
Jet Propulsion Laboratory, 4800 Oak Grove Drive, Pasadena, CA 91109, USA
\item
Instituto de Astrof\'isica de Canarias (IAC), C/v{\'\i}a L{\'a}ctea S/N, 38200 La Laguna, Spain 
\item
Departamento de Astrof\'isica, Universidad de La Laguna, Spain 
\item
Dipartimento di Astronomia, Universit\`a di Bologna, via Ranzani 1, 40127 Bologna, Italy
\item
Center for Astrophysics and Space Astronomy, 389 UCB, University of Colorado, Boulder, CO 80309, USA
\item
Institute for Astronomy, University of Edinburgh, Royal Observatory, Blackford Hill, Edinburgh EH9 3HJ, UK
\item
Department of Physics, University of Oxford, Keble Road, Oxford OX1 3RH, UK
\item
School of Physics and Astronomy, The Raymond and Beverly Sackler Faculty of Exact Sciences, Tel Aviv University, Tel Aviv 69978, Israel
\item
NASA Ames, Moffett Field, CA 94035, USA
\item
Dipartimento di Astronomia, Universita di Padova, Vicolo dell'Osservatorio 3, I-35122, Italy
\item
Department of Physics \& Astronomy, University of British Columbia, 6224 Agricultural Road, Vancouver, BC V6T 1Z1, Canada
\end{enumerate}

\end{appendix}
\end{document}